\documentclass[twocolumn,tighten]{aastex63}
\usepackage{amsmath,amssymb,bm}
\usepackage{times}
\usepackage{colordvi}

\received{2018 February 12} \revised{2021 January 4} \accepted{2021 January 5} \published{}
\reportnum{~\hfill \textit{final revision}} %, Jan 4, 2021}  % <- AASTeX6.2
\shortauthors{Kliem et al.}
\shorttitle{Flux Rope Formation Preceding a CME} 

 \newcommand{\B}[1]{{#1}}   

\begin{document}

\title{Non-equilibrium Flux Rope Formation by Confined Flares Preceding a Solar Coronal Mass Ejection}

\author[0000-0002-5740-8803]{Bernhard Kliem}
\affiliation{Institute of Physics and Astronomy, University of Potsdam, Potsdam 14476, Germany}

\author[0000-0002-5865-7924]{Jeongwoo Lee}
\affiliation{Institute for Space Weather Sciences, New Jersey Institute of Technology, University Heights, Newark, NJ 07102, USA}

\author[0000-0003-4618-4979]{Rui Liu}
\affiliation{Department of Geophysics and Planetary Sciences, University of Science and Technology of China, Hefei, 230026, China}

\author[0000-0002-8574-8629]{Stephen M. White}
\affiliation{Air Force Research Laboratory, Space Vehicles Directorate, Albuquerque, NM 87117, USA}

\author[0000-0002-6178-7471]{Chang Liu}
\affiliation{Institute for Space Weather Sciences, New Jersey Institute of Technology, University Heights, Newark, NJ 07102, USA}

\author{Satoshi Masuda}
\affiliation{Institute for Space-Earth Environmental Research, Nagoya University, Aichi 464-8601, Japan}

\correspondingauthor{Bernhard Kliem}
\email{bkliem@uni-potsdam.de}

\begin{abstract} 
We present evidence that a magnetic flux rope was formed before a coronal mass ejection (CME) and its associated long-duration flare during a pair of preceding confined eruptions and associated impulsive flares in a compound event in NOAA Active Region 12371. Extreme-ultraviolet images and the extrapolated nonlinear force-free field show that the first two, impulsive flares, SOL2015-06-21T01:42, result from the confined eruption of highly sheared low-lying flux, presumably a seed flux rope. The eruption spawns a vertical current sheet, where magnetic reconnection creates flare ribbons and loops, a nonthermal microwave source, and a sigmoidal hot channel which can only be interpreted as a magnetic flux rope. Until the subsequent long-duration flare, SOL2015-06-21T02:36, the sigmoid's elbows expand, while its center remains stationary, suggesting non-equilibrium but not yet instability. The ``flare reconnection'' during the confined eruptions acts like ``tether-cutting reconnection'' whose flux feeding of the rope leads to instability. The subsequent full eruption is seen as an accelerated rise of the entire hot channel, seamlessly evolving into the fast halo CME. Both the confined and ejective eruptions are consistent with the onset of the torus instability in the dipped decay index profile which results from the region's two-scale magnetic structure. We suggest that the formation or enhancement of a non-equilibrium but stable flux rope by confined eruptions is a generic process occurring prior to many CMEs.\\ 
\end{abstract}  % 241 words

% \keywords{% Solar active regions, Solar filaments, 
%           Active solar corona, 
%           Solar coronal mass ejections, Solar flares, 
%           Solar magnetic fields, Solar magnetic reconnection}

\section{Introduction}\label{s:intro}

A magnetic flux rope is a fundamental structure of magnetic field twisted about an axis field line. It is subject to torus  or kink instability to erupt, which is recognized as important driver of solar and heliospheric disturbances and thus an important factor in space weather. It has been suggested that flux ropes can occur in the solar atmosphere either as the direct result of an emerging flux rope which partly keeps its coherence in the emergence process \citep{LowBC1996} or from a sheared coronal magnetic arcade via the magnetic reconnection associated with a photospheric flux-cancellation process \citep{vanBallegooijen&Martens1989}, often referred to as ``tether-cutting reconnection'' \citep{Moore&Roumeliotis1992}. A transition from arcade to flux rope structure, involving reconnection, occurs in the former case as well, as the upper half of an emerging flux rope is an arcade and reconnection within the lower half is required to release the entrained mass and allow the lower half to emerge. The formation time of flux ropes in the corona relative to the onset time of eruptions is a key issue whose resolution is likely to decide between the two categories of models for solar eruptions \cite[e.g.,][]{Forbes2000, ChenPF2011}. Ideal MHD models \citep{vanTend&Kuperus1978} require a preexisting flux rope whose equilibrium turns unstable; the instability then drives reconnection. Alternative to such onset and driving, reconnection models \citep{Moore&al2001, Karpen&al2012} require self-amplifying reconnection in a magnetic arcade, producing a flux rope during the eruption. 

The detection of flux ropes in the low corona, where eruptions typically originate, and prior to the onset of eruptions, has been elusive. Twisted fine structures are often seen when eruptions are underway, both for rising filaments/prominences \cite[e.g.,][]{Vrsnak&al1991, Romano&al2003, Koleva&al2012, LiT&ZhangJ2015, XueZ&al2016} and coronal mass ejections (CMEs) \citep{Dere&al1999}, but are difficult to observe unambiguously prior to eruption onset. The existence of so-called bald-patch sections of the source region's polarity inversion line (PIL), where the horizontal component of the magnetogram points in the inverse direction, indicates that a transition from arcade to flux rope structure has at least begun, especially for bipolar regions with relatively straight PILs. This is subject to the correct resolution of the $180^\circ$ ambiguity of the horizontal field component and has not yet been investigated systematically. Flux ropes are increasingly also indicated by magnetogram extrapolations \cite[e.g.,][]{Canou&al2009, YellesChaouche&al2012, GuoY&al2013, ZhaoJ&al2016, LiuR&al2016, Green&al2017, James&al2018, DuanA&al2019}, but the results of this technique still depend strongly on the extrapolation scheme and code employed (compare, e.g., \citealt{JingJ&al2018} and \citealt{DuanA&al2019}). 

Many authors suggest that solar filaments/prominences form in flux ropes \citep{Rust&Kumar1994, LowBC1996, Kumar&al2011}. Strong observational support for the conjecture has been found for quiescent prominences \cite[e.g.,][]{Leroy1989, LiX&al2012, Bak-Steslicka&al2013}, however, it is not clear whether active-region filaments form in flux ropes or in sheared arcades \citep{Mackay&al2010, Patsourakos&al2020}. This is illustrated by a recent observation of a confined filament eruption with the highest resolution currently available. Although the extrapolation of a pre-eruption vector magnetogram indicated the existence of a flux rope, observational signatures of twist became unambiguous only in the course of the eruption \citep{WangH&al2015}. 

Sigmoidal structures often seen in soft X-ray and extreme-ultraviolet (EUV) coronal images are suggestive of flux ropes \citep{Rust&Kumar1996, Titov&Demoulin1999, McKenzie&Canfield2008, Savcheva&vanBallegooijen2009}, but it has also been suggested that they indicate a highly sheared arcade \citep{Antiochos&al1994, DeVore&Antiochos2000, Moore&al2001}. It is quite likely that sigmoids consisting of a collection of J shaped loops outline a highly sheared arcade, whereas a dominant single, continuous S shape indicates the formation of a flux rope \citep{Green&Kliem2009, Green&al2011}. In the soft X-ray range, where sigmoids tend to be visible for several days, the continuous S shape emerges through a transition from a double J shape, consistent with flux rope formation from an arcade through reconnection. In the EUV, continuous S shaped sigmoids have been observed in lines formed at and above 7~MK and were termed ``hot channel''; these structures are also visible at the limb, where they form rising arcs when seen from the side or blobs when seen along the axis \citep{LiuR&al2010, ChengX&al2011, ZhangJ&al2012, ChengX&al2014}. Hot channels tend to appear in close temporal association with flaring activity, mostly only during flares, due to the high plasma temperatures involved. It is still unclear whether the underlying magnetic structure typically forms with the hot channel or earlier. A couple of cases of repeated illumination by hot plasma exist and indicate that the magnetic structure was formed at or before the first illumination, but it is not yet clear whether this is typical or rare. Continuous S shaped sigmoidal structures on disk and hot channels above the limb currently appear to be the most direct indication for the existence of a flux rope in active regions. However, their timing relative to the onset of an eruption is still an open question.

Only a minority of soft X-ray sigmoids show the formation of a continuous S shape. In a first systematic study of this phenomenon, comprising six cases, \citet{Green&Kliem2014} found their formation time to lie in the range 4--14~hr prior to the onset of an eruption. Of the several hundred hot channels detected in recent years \citep{ZhangJ&al2015, Nindos&al2015, ChengX&DingMD2016b}, we are aware of only $\gtrsim\!10$ cases shown to be formed prior to eruption onset. Most of them occurred during small-scale precursor activity or during a slow-rise phase preceding eruption onset, with the lead times lying between 6 and $\sim$140~min \citep{LiuR&al2010, ZhangJ&al2012, ChengX&al2013, ChengX&al2014, ChengX&al2015, ChenB&al2014, JoshiB&al2017}. 
% LiuR et al.   2010: 50 min onset of slow-rise phase
% ZhangJ et al. 2012:  6 min in slow-rise phase
% ChenB et al.  2014: 42 min in slow-rise phase
% ChengX et al. 2013: 95 min & 13 min, both in slow-rise phase
% ChengX et al. 2014: 14 min in slow-rise phase
% ChengX et al. 2015: 4:35 hr in prec. conf. flare (18-Apr-2014); 105 min intermittent in slow-rise phase (10-Sep-2014)
% JoshiB et al. 2017: 140 min in small-scale brightening activity/(slow-rise phase(?))
% James et al. 2017, 2018: ~1 hr in C5, then ~1 hr gap; Note: WangW+19 find no gap
% WangW et al. 2019: 5 hr incl. 4 C flares continuous visibility (same event as James+!)
% LiuLJ et al. 2018: X2 3 hr earlier, HC continuously visible, but other, compact struct erupts!
Of special interest are the six cases of hot channel detection during a confined flare (not associated with a CME) which preceded an eruptive flare (associated with a CME) by several hours. In these events, a hot channel appeared or brightened during the confined flare(s) and brightened again in identical position and shape at the onset of the eruptive flare. It remained continuously visible in the 7~hr interval between the limb events studied by \citet{Patsourakos&al2013} and possibly also in the 5~hr interval (with four C-class flares) prior to the event on disk studied by \citet{WangW&al2019}. However, \citet{James&al2017} concluded for the latter event that the hot channel appeared in the final and strongest preceding confined flare and reappeared two hours later in the eruptive flare. The hot channels studied by \citet{ChengX&al2015} and \citet{Chintzoglou&al2015} disappeared in the 4.5~hr and 12~hr intervals, respectively, between the initial confined and subsequent eruptive flare in these disk events. In the case of hot channel formation by the confined X2.2 flare preceding the eruptive X9.3 flare on 2017 Sep~9 \citep{LiuLJ&al2018}, there are indications that it was another, more compact structure which erupted in the X9.3 flare. 

In this paper, we report on unambiguous observational evidence for the presence of a flux rope in a hot channel during an impulsive, confined flare that occurred in a compound event with another confined and a subsequent eruptive flare. Moreover, for the first time, an event exhibits a well defined sigmoidal hot channel on disk continuously between two flares, so that its evolution can clearly be followed from formation in one event to eruption in the other. This provides unambiguous evidence for the interpretation of the hot channel as a flux rope and detailed information about its genesis. Also, it stimulates a consideration of the relevance of confined flares for flux rope formation compared to slow tether-cutting reconnection driven from the photosphere, which is important for the forecast of CMEs. The specifics of the eruptive flare were already analyzed in \citet{LeeJ&al2018}. The compound event was recently also analyzed by \citet{Mitra&al2020}; we will discuss their partly different conclusions below.

\section{Observations}\label{s:obs}

\subsection{Flare Light Curves and Ribbons}\label{ss:lightcurves}

%~~~~~~~~~~~~~~~~~~~~~~~~~~~~~~~~~~~~~~~~~~~~~~~~~~~~~~~~~~~~~~~~~~~~~~~~~~~~~~~
\begin{figure}[t]                                                          % f01
\centering
\includegraphics[width=\linewidth]{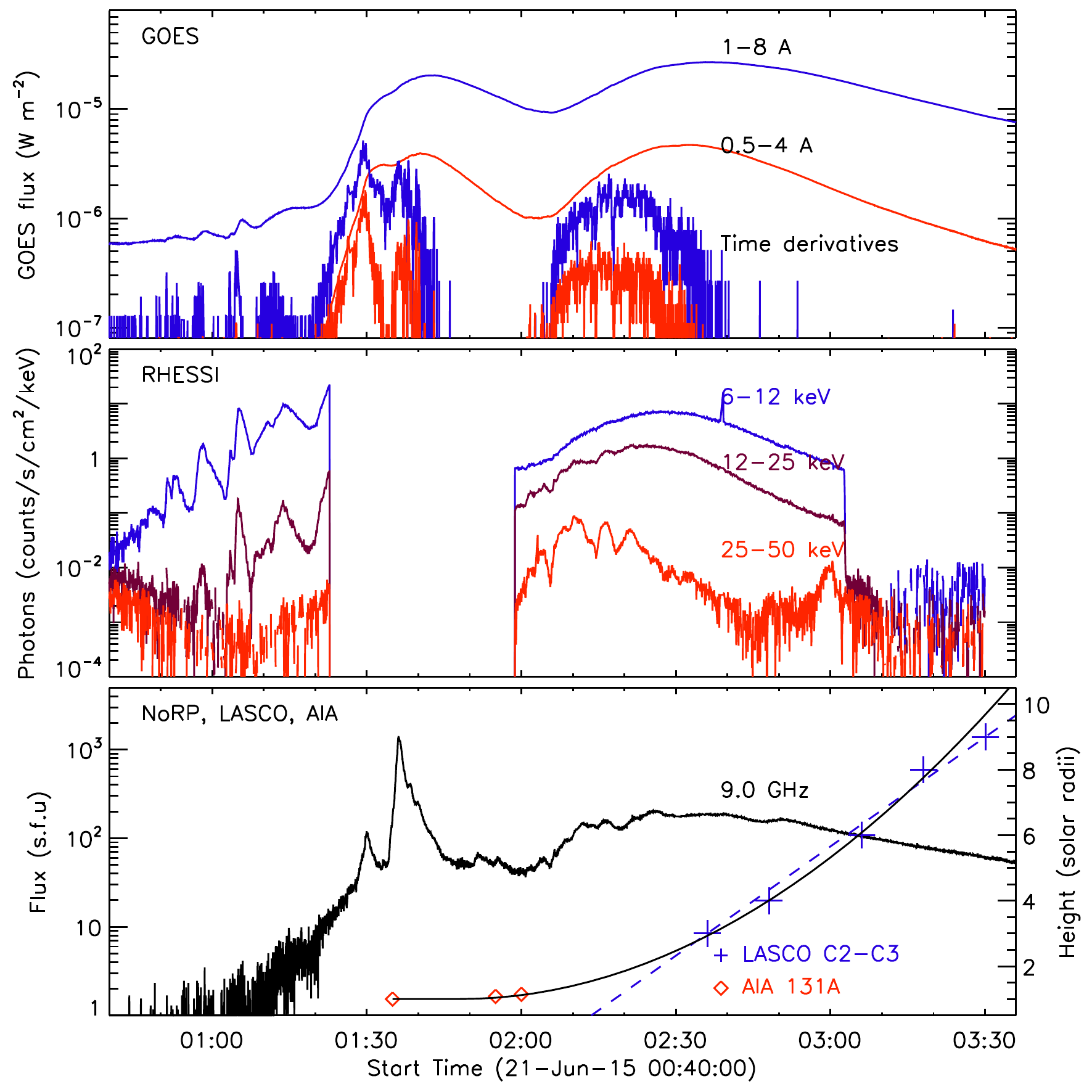} %  <-- 2-column style
\caption{
 (top) \textsl{GOES} soft X-ray light curves of the precursors, two impulsive flares ($\approx$M1.2 and M2.1, jointly listed as SOL2015-06-21T01:42), and long-duration flare (M2.6, SOL2015-06-21T02:36), and the time derivatives of the light curves. 
 (middle) \textsl{RHESSI} hard X-ray light curves. 
 (bottom) NoRP 9~GHz flux, LASCO CME height data, and AIA position data of the second (main) hot channel along slit S1 in Figure~\ref{f:td131}. The dashed line is a linear fit to the LASCO data and the solid line is a quadratic fit to the combined AIA and LASCO data.} 
\label{f:lightcurves}
\end{figure}
%~~~~~~~~~~~~~~~~~~~~~~~~~~~~~~~~~~~~~~~~~~~~~~~~~~~~~~~~~~~~~~~~~~~~~~~~~~~~~~~

%~~~~~~~~~~~~~~~~~~~~~~~~~~~~~~~~~~~~~~~~~~~~~~~~~~~~~~~~~~~~~~~~~~~~~~~~~~~~~~~
\begin{figure*}[t!]                                                        % f02
\centering
\includegraphics[width=.95\linewidth]{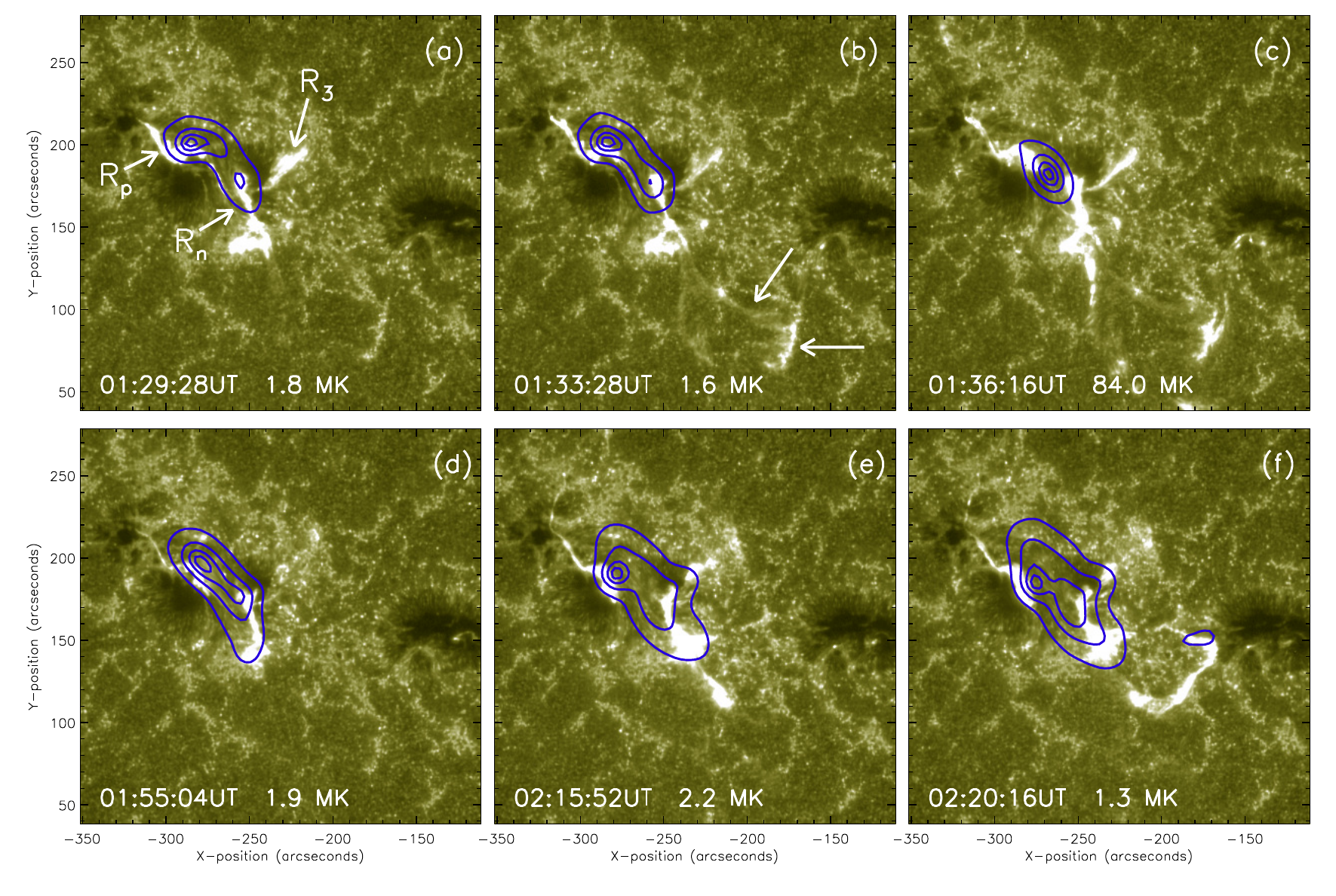} 
\caption{Evolution of flare ribbons in AIA 1600~{\AA} images.
 (a) First impulsive flare; 
 (b) extension along the PIL during the eruption of filamentary material in the second impulsive flare (marked by arrows); 
 (c) peak of second impulsive flare;
 (d) onset time of long-duration flare;
 (e) ribbon extension and 
 (f) nearly fully closed triangular ribbon, both during the rise of the long-duration flare. 
The blue contours show the 17~GHz radio source, whose peak brightness temperature (at 12\arcsec\ resolution) is also listed. The main ribbon pair is marked in (a) as R$_\mathrm{p}$ and R$_\mathrm{n}$, and a secondary ribbon is marked as R$_3$.} 
\label{f:ribbons}
\end{figure*}
%~~~~~~~~~~~~~~~~~~~~~~~~~~~~~~~~~~~~~~~~~~~~~~~~~~~~~~~~~~~~~~~~~~~~~~~~~~~~~~~

The compound event consists of a series of small precursors of increasing amplitude, with the final one peaking at $\approx$01:14~UT slightly above the \textsl{GOES} C1 level, two strongly overlapping impulsive flares peaking, respectively, at $\approx$01:33 and 01:42~UT and $\approx$M1.2 and M2.1 level, which are jointly listed as SOL2015-06-21T01:42, and a more gradual long-duration M2.6 flare, SOL2015-06-21T02:36 (Figure~\ref{f:lightcurves}). The peak level of the first impulsive flare in the standard 1--8~{\AA} band is taken at the well defined peak time of its 0.5--4~{\AA} light curve. The long-duration flare has a rather slow rise of $\sim$40~min duration, which can still be considered to be an impulsive phase because it comprises fluctuating nonthermal emissions and the higher-energy (25--50~keV) hard X-rays are well synchronized with the rise phase of the soft X-ray flux, as in the classical picture of an impulsive flare phase. Figure~\ref{f:lightcurves} displays the soft X-ray light curves from the \textsl{Geostationary Operational Environmental Satellite (GOES)} and the hard X-ray light curves from the \textsl{Reuven Ramaty High Energy Solar Spectroscopic Imager (RHESSI)} \citep{LinR&al2002}. Unfortunately, \textsl{RHESSI} missed the impulsive flares, which were likely stronger than the long-duration flare in the 6--25~keV range based on the stronger 6--12~keV and microwave fluxes. The 9~GHz microwave flux from the Nobeyama Radio Polarimeters (NoRP) provides information about nonthermal energy release during \textsl{RHESSI} night. This emission is dominantly nonthermal, because the thermal bremsstrahlung emission at 9~GHz, which can be calculated from the \textsl{GOES} soft X-ray flux under the assumption of optically-thin bremsstrahlung using the temperature and emission measure fits to the \textsl{GOES} data adopting coronal abundances \citep{White&al2005}, stays below 10~sfu during the impulsive flares, but the flares reach 100~sfu and nearly 1400~sfu. The two impulsive flares can here clearly be distinguished. Their peaks at 01:30 and 01:36~UT coincide with the rise of the corresponding soft X-ray emission phases. A comprehensive analysis of the microwave emission is given in \citet{LeeJ&al2017, LeeJ&al2018}. The compound flare event is accompanied by a fast halo CME which reaches a velocity of $\approx$1300~km\,s$^{-1}$ in the data of the \textsl{SOHO}/LASCO C2 and C3 telescopes \citep{Brueckner&al1995}, which are included in Figure~\ref{f:lightcurves}. 

The source active region, NOAA 12371, is located near disk center (approximately N10E17) and composed of a bipolar spot group in the east and a separate unipolar spot in the west, which is of the same (negative) polarity as the adjacent leading spot of the group (Figure~\ref{f:ribbons}(a); see Section~\ref{s:magnetic} for the magnetogram). The flares and their final precursor share a common pair of main ribbons which begin to brighten at the PIL within the bipolar spot group already from $\approx$01:10~UT and then evolve continuously through the compound event. The formation of this ribbon pair implies that all four phases of the event trigger magnetic reconnection in a vertical (``flare'') current sheet and involve the release of energy stored in the coronal field above this section of the PIL. The strong displacement of the ribbons along the PIL indicates that the region's core flux above the PIL in the spot group is highly sheared in left-handed sense. Figure~\ref{f:ribbons} displays the ribbons and sunspots in UV 1600 {\AA} images from the Atmospheric Imaging Assembly \cite[AIA,][]{Lemen&al2012} on board the \textsl{Solar Dynamics Observatory (SDO)} at six time intervals during the flares, overlaid with Nobeyama Radioheliograph (NoRH) 17~GHz maps. 

%~~~~~~~~~~~~~~~~~~~~~~~~~~~~~~~~~~~~~~~~~~~~~~~~~~~~~~~~~~~~~~~~~~~~~~~~~~~~~~~
\begin{figure*}[t!]                                                        % f03
\centering
\includegraphics[width=.75\linewidth]{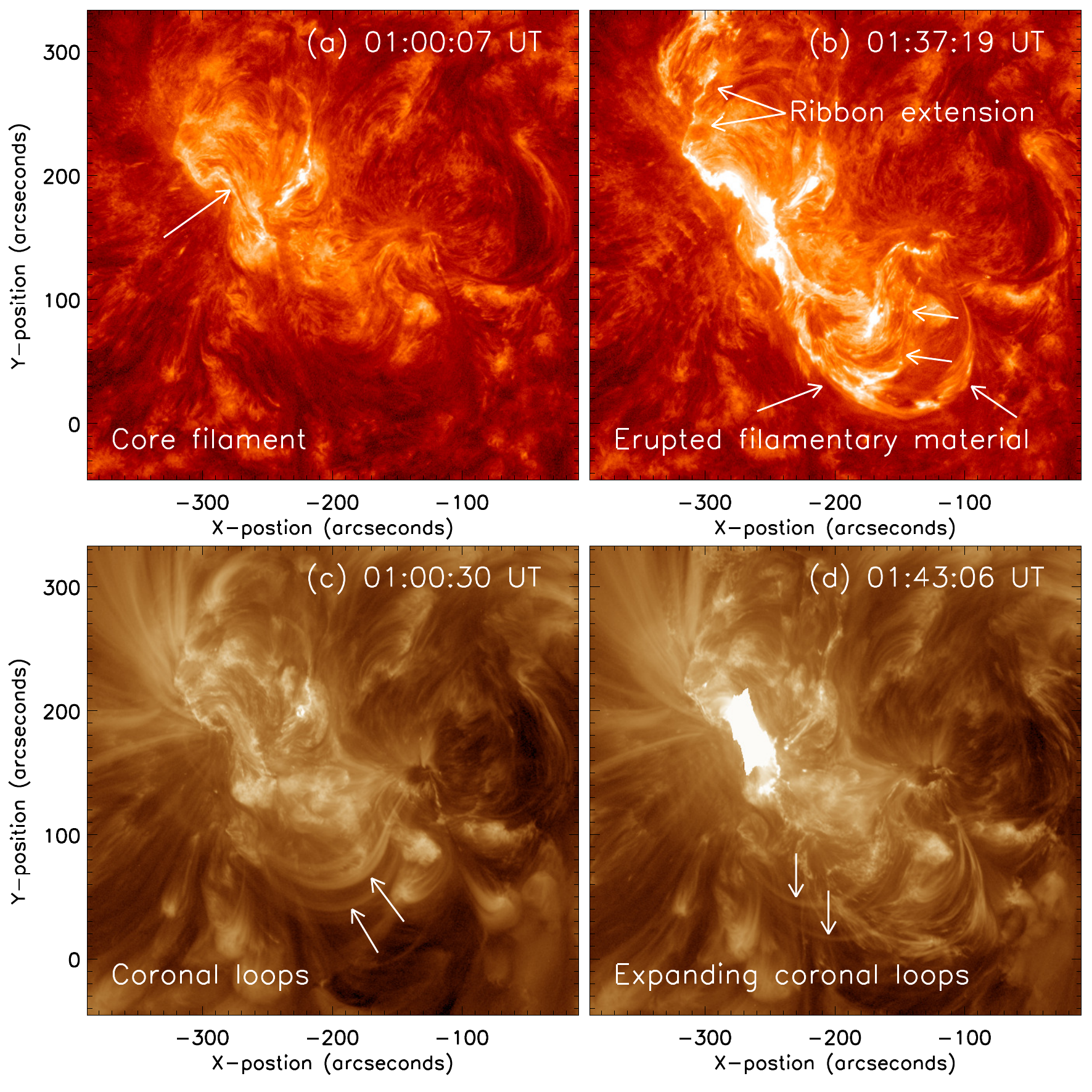} 
\caption{
 (a--b) AIA 304~{\AA} images showing the narrow filament in the core of the active region, which does not erupt, and the filamentary material which erupts in the second impulsive flare and drains to the unipolar spot along two arcs, each of which is marked by two arrows. The northern ribbon extension is also marked. 
 (c--d) AIA 193~{\AA} images showing coronal loops which are nearly cospatial with the southern elbow of the sigmoidal main hot channel when it first lights up and with the initially overlying, moderately sheared flux (orange field lines in Figure~\ref{f:flines}). Their expansion is commensurate with the expansion of the hot channel's southern elbow. 
(An animation of this figure is available.)} 
\label{f:aia304+193}
\end{figure*}
%~~~~~~~~~~~~~~~~~~~~~~~~~~~~~~~~~~~~~~~~~~~~~~~~~~~~~~~~~~~~~~~~~~~~~~~~~~~~~~~

Figure~\ref{f:ribbons}(a) shows the well-developed main ribbon pair, marked as R$_\mathrm{p}$ and R$_\mathrm{n}$ (indicating polarity), near the peak time of the nonthermal microwave enhancement in the first impulsive flare, with their brightest parts lying in the same position as the ribbons of the final precursor. The nonthermal 17~GHz emission originates between the ribbons, where particle acceleration in a vertical current sheet is expected. The formation of a vertical current sheet is also indicated by the southward extension of the R$_\mathrm{p}$ ribbon (Figures~\ref{f:ribbons}(a)--(b)), which yields a section of parallel ribbon formation in the core of the spot group analogous to a classical two-ribbon flare. The ribbons brighten throughout the impulsive flares and weaken somewhat subsequently, until they brighten again in the long-duration flare. Up to the onset of the long-duration flare, they show nearly no separation from the PIL, which indicates that both impulsive flares are associated with a confined eruption, as usual. This is confirmed by the stationary, compact character of the microwave source throughout these flares \citep{LeeJ&al2018}. AIA 304~{\AA} images show that a narrow filament, extending along the PIL between the ribbons (``core filament'' in Figure~\ref{f:aia304+193}(a)), does not erupt. Rather, it is gradually masked by the increasing brightness around it without moving and without showing any significant internal brightening. This signature of a partial eruption (involving only flux above the filament) is consistent with the active region's magnetic structure, analyzed in Section~\ref{s:magnetic}. 

During the second impulsive flare, the ribbon in the negative polarity shows a long, transient extension along the PIL in southwestward direction; this weak, quickly fading extension is marked by arrows in Figure~\ref{f:ribbons}(b). The ribbon in the positive polarity then also brightens and extends northward (this is only very weakly indicated in the 1600~{\AA} image in Figure~\ref{f:ribbons}(c) but clear in the 304~{\AA} images, see Figure~\ref{f:aia304+193}(b)). Thus, magnetic flux has likely risen above these extended ribbons, which is also suggested by the fact that filamentary material now drains to the unipolar spot along two arcs in the area of the southwestward ribbon extension (also see Figure~\ref{f:aia304+193}(b) and the accompanying animation of the AIA 304~{\AA} data). These arcs remain confined during the second impulsive flare (but the outer, southern arc indicates an impulsive southward expansion from the onset of the long-duration flare). H$\alpha$ data from the GONG network show that the major part of a filament in the area of the southwestward ribbon extension disappears during $\sim$01:35--01:45~UT and reforms only $\sim$10~hours later. 
% ftp://gong2.nso.edu/HA/hag/201506/20150621/
The weak ribbon extensions during the second impulsive flare indicate that more extended magnetic flux rises in this phase, while the flare energy release (the reconnection) remains strongest in the core of the active region (in the area of the first impulsive flare), although no further eruption of flux is apparent in this region. This unexpected behavior is put in a coherent picture by the data of the hot channel and the magnetic structure of the active region in the subsequent sections. 

With the rise of the long-duration flare, both ribbons intensify and separate clearly. The ribbon in the negative polarity also lengthens toward the unipolar spot. Additionally, this ribbon nearly closes around a parasitic polarity (Figure~\ref{f:ribbons}(e)--(f)). This aspect has been studied in detail in \citet{LeeJ&al2018} and is not relevant here. 

A third ribbon, marked as R$_3$ in Figure~\ref{f:ribbons}(a), brightens from $\approx$00:50~UT, simultaneous with the first precursor, and remains visible throughout the event, with variable brightness. \citet{Mitra&al2020} have shown that this ribbon traces a quasi-separatrix layer (QSL) between the flux overlying the core filament and flux that closes locally above weak mixed polarity in the area of the third ribbon, where another filament lies. This filament slowly rises and is activated during the precursors, which initiates the third ribbon, but it does not erupt. From the timing of the third ribbon, \citet{Mitra&al2020} suggested that the underlying reconnection triggered the whole compound event. Reconnection in the adjacent QSL can weaken the flux overlying the core filament, perturbing the force balance in the filament channel, and indeed provide a trigger of its destabilization. However, the main energy storage and release of the compound event occurs at the ``main'' section of the PIL (along the core filament), where the primary ribbon pair, the flare loops, and the nonthermal microwave source form. The total vertical current flowing at the photopspheric level in the area of the third ribbon is more than an order of magnitude weaker than the total current in the area of the primary ribbon pair \citep{Mitra&al2020}. Therefore, we focus on this core region in analyzing and interpreting the event.

\subsection{Sigmoidal Hot Channel and CME}\label{ss:HC}

%~~~~~~~~~~~~~~~~~~~~~~~~~~~~~~~~~~~~~~~~~~~~~~~~~~~~~~~~~~~~~~~~~~~~~~~~~~~~~~~
\begin{figure*}[t!]                                                        % f04
\centering
\includegraphics[width=.95\linewidth]{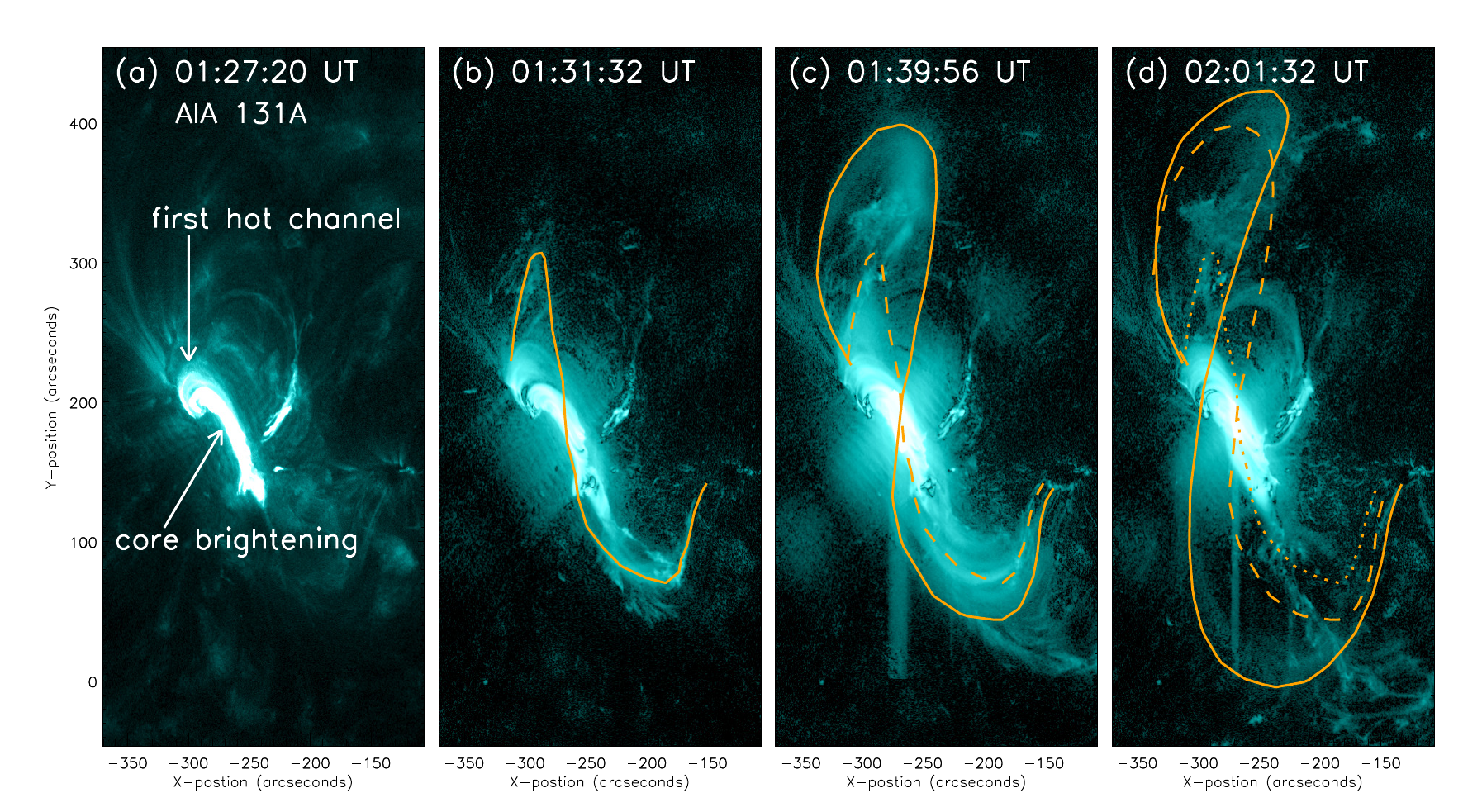}
\caption{Evolution of hot plasma showing (a) the structure of the final precursor and first 
impulsive flare and (b)--(d) the strongly sigmoidal second (main) hot channel in AIA 131~{\AA} images. The solid yellow lines in (b)--(d) outline the second hot channel at the time of each frame, and the dashed and dotted yellow lines are copied from the previous frames for comparison. 
(An animation of this figure is available.)}
\label{f:sigmoid}
\end{figure*}
%~~~~~~~~~~~~~~~~~~~~~~~~~~~~~~~~~~~~~~~~~~~~~~~~~~~~~~~~~~~~~~~~~~~~~~~~~~~~~~~

This event forms two sigmoidal hot channels which outline the erupting flux. In Figure~\ref{f:sigmoid}, we use AIA 131~{\AA} images to show the morphology of hot plasma during the final precursor and first impulsive flare, which includes a first hot channel, and three snapshots of the second, main hot channel, which this investigation focuses on. Figure~\ref{f:sigmoid}(a) shows the hot plasma during the rise of the first impulsive flare as an elongated, slightly sigmoidal, stationary source that follows the PIL in the core of the sunspot group and extends nearly exactly between the brightest parts of the flare ribbons. This source consists of a bright core (marked in Figure~\ref{f:sigmoid}(a)) and a weaker, diffuse envelope. The bright core is seen with nearly identical shape during the final precursor (narrower and slightly shorter) and throughout the two impulsive flares (slightly growing in length and width). From its spatio-temporal synchronization with the ribbons, it is clear that it represents the first flare loops, which grow seamlessly from the final precursor through the impulsive flares. The formation of flare loops between the pair of main ribbons R$_\mathrm{p}$ and R$_\mathrm{n}$ demonstrates that reconnection in a vertical current sheet acted in the final precursor and confined flares. The diffuse envelope appears during the first impulsive flare. We interpret this structure as the signature of the flux erupting in this phase, i.e., as a first hot channel in the event. From the similarity of the flare ribbons and loops, it is very likely that the flux erupting in the final precursor is of similar geometry as the flux erupting in the first impulsive flare, but does not release enough energy to be illuminated by plasma emitting at 131~{\AA}. 

A second, classical hot channel develops near the peak of the first impulsive flare, which is shortly after the estimated onset time of the second impulsive flare if both flares have similar rise times (see Figure~\ref{f:lightcurves}(a)). Its southern part begins to brighten at $\approx$01:28:30~UT and is clearly seen in full extent at $\approx$01:30~UT. From $\approx$01:29:30~UT the channel also develops a northern elbow, thus showing a clear sigmoidal shape, which is typical for such structures on disk but particularly strongly exhibited in the present event. The reverse S shape indicates negative helicity, similar to the first hot channel and flare loops. 

The second hot channel does not form through a lengthening of the 131~{\AA} core source or out of the first hot channel, but is a new structure whose middle part crosses these structures with a small angle. Individual loops of increasing size can be seen to brighten successively in each elbow of the sigmoid from $\approx$01:28:30~UT (see the animation of Figure~\ref{f:sigmoid}). These merge into the single sigmoidal structure of the second hot channel during $\approx$01:40--01:45~UT. Its foot point in the negative polarity lies in the unipolar spot, doubtlessly very far from the foot point area of the flux that erupted to produce the bright 131~{\AA} core source and the flare ribbons up to this point. The conjugate foot point is more difficult to discern, due to complex background structures, but also appears to lie separate from the foot point of the initially erupting flux indicated by the first hot channel. The foot points given in Figure~\ref{f:sigmoid}(b)--(d) were determined from the appearance of the second hot channel in AIA 131~{\AA} images; they are perfectly consistent with the extent of the fully developed flare loop arcade in AIA images of the main and decay phases of the long-duration flare. The flare loop arcade then extends exactly between these foot point locations (see the animation of Figure~\ref{f:sigmoid}). 

Both elbows of the newly formed second hot channel immediately begin to expand (Figure~\ref{f:sigmoid}(c)). However, until $\approx$01:55~UT, the center of the sigmoid remains nearly stationary at the position of the bright 131~{\AA} core source, which is unlikely to be a projection effect because the elbows expand strongly and asymmetrically. Thus, the second impulsive flare is characterized by the strong expansion of the elbows of the second hot channel, which initially are nearly cospatial with the ribbon extensions discussed above. The end points of the channel also remain fixed during this expansion, so that it takes a very strong sigmoidal shape which can only be explained as the signature of a flux rope \citep{Rust&Kumar1996} or of a flux bundle in the surface of a flux rope \citep{Titov&Demoulin1999}. 

%~~~~~~~~~~~~~~~~~~~~~~~~~~~~~~~~~~~~~~~~~~~~~~~~~~~~~~~~~~~~~~~~~~~~~~~~~~~~~~~
\begin{figure*}[t!]                                                        % f05
\centering 
\includegraphics[width=.75\linewidth]{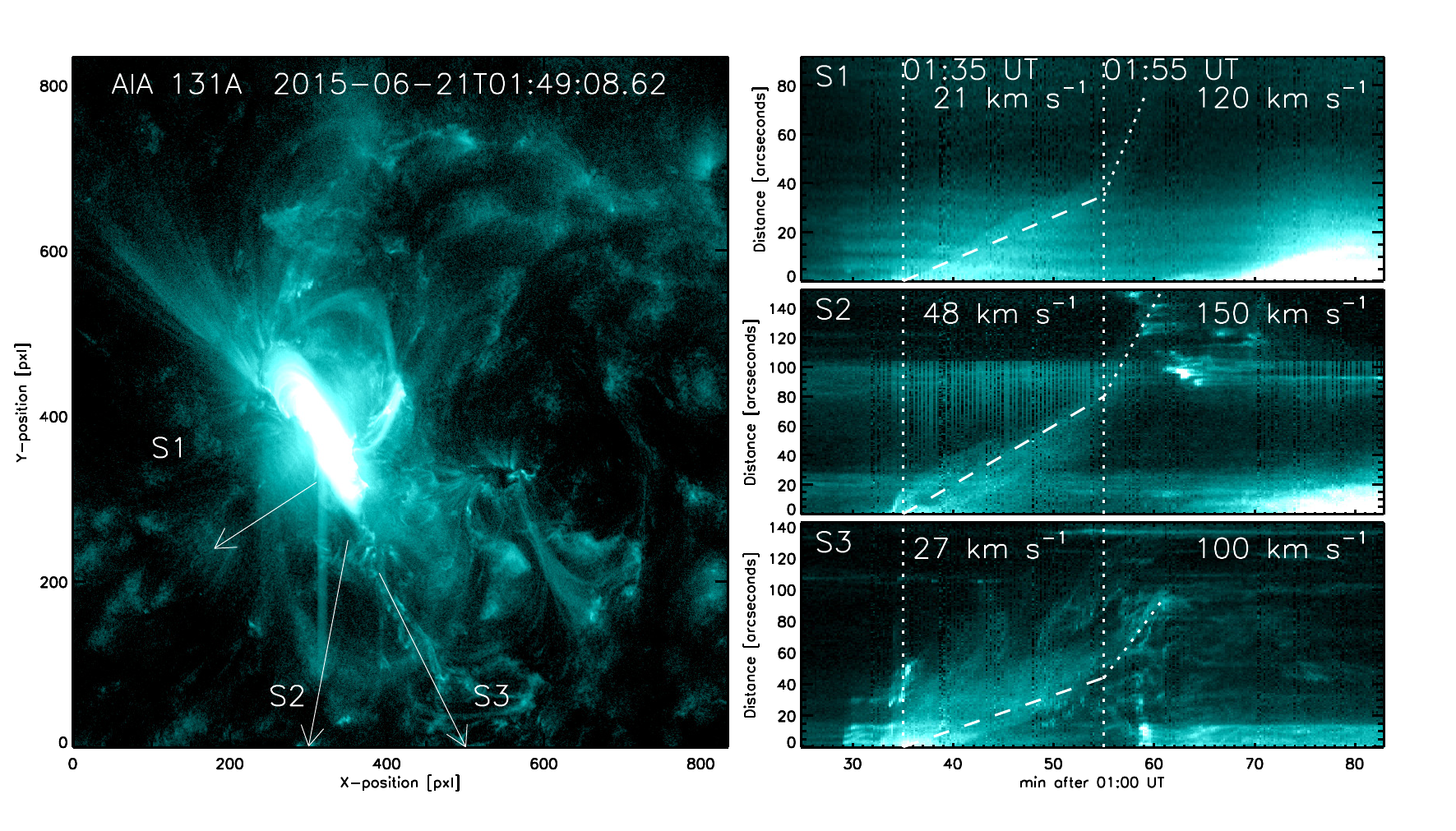}
\caption{Time-distance diagrams for the motion of the strongly sigmoidal second hot channel seen in AIA 131~\AA\ images along the slits S1--S3.}
\label{f:td131}
\end{figure*}
%~~~~~~~~~~~~~~~~~~~~~~~~~~~~~~~~~~~~~~~~~~~~~~~~~~~~~~~~~~~~~~~~~~~~~~~~~~~~~~~

The alternative interpretation as a sheared arcade that leans over the PIL at both ends to form a sigmoid \citep{Antiochos&al1994, DeVore&Antiochos2000} can be ruled out. When such a configuration is formed, the sigmoidal shape is only weakly indicated, in particular, the ends of the sigmoid do not turn around sufficiently to point toward the middle of the structure. This feature of a strongly sigmoidal shape is \emph{natural} for flux winding in a flux rope of sufficient twist (of $\approx$1~turn or higher). It can occur in a sheared arcade if two conditions are met: if the arcade rises strongly (far stronger than in \citealt{Antiochos&al1994} and \citealt{DeVore&Antiochos2000}), such that its leg sections expand along the PIL beyond the fixed end points, and if it is seen in nearly vertical projection. However, two properties of the second hot channel in the present event are at variance with this scenario. First, the channel shows a very strong sigmoidal shape from the beginning (Figure~\ref{f:sigmoid}(b)), before any strong expansion commences. Second, the rise of an arcade also involves a strong expansion of its cross section \cite[see, e.g., the outermost field line of the rising arcade in Figure 3(a,b) of][]{Karpen&al2012}, and this is not shown by the second hot channel in the present event. 

Since the elbows of the second hot channel immediately begin to expand, the underlying flux rope is in a non-equilibrium state. This is not surprising because the rope is formed in a dynamical manner and further exposed to the outflow of the fast reconnection in the vertical current sheet of the impulsive flares. Flux in the upward reconnection outflow adds to the flux erupted in the precursor and first impulsive flare. It is frozen in the hot plasma that forms the second hot channel \cite[see, e.g., Figure~1 in][]{LinJ&al2004}. In this picture, the stationarity of the channel's middle section is consistent with the confined nature of the impulsive flares. This section of the flux rope is held down by overlying flux in the center of the bipolar spot group. The picture also suggests that the newly added flux crosses the PIL in projection and thus, as observed, makes an angle with the first, highly sheared flare loops and the first hot channel, which follow the PIL closely and actually cross it in the opposite direction.

Finally, the middle section of the second hot channel begins an accelerated eastward motion around 01:55~UT (see Figure~\ref{f:sigmoid}(d)), during which the channel rapidly fades; its southern elbow then also expands faster. These features are consistent with a primarily upward motion from the eastern source location. The motion can be followed in the animated 131~{\AA} image sequence (Figure~\ref{f:sigmoid}) for $\sim$100~arcsec until $\approx$02:08~UT. We trace the position of the second hot channel in 131~\AA\ images along the three lines shown in Figure \ref{f:td131} (``slits'' S1--S3), pointing roughly in the southeast, south, and southwest directions; background structures make it more difficult to trace the northward expansion. The two vertical dotted lines in the figure mark the approximate start time of the first expansion ($\approx$01:35~UT) and that of the second expansion ($\approx$01:55~UT). The first expansion yields projected speeds of 21--48~km\,s$^{-1}$ along the three slits. Note that S1 is placed slightly south of the stationary sigmoid center, to avoid a region of enhanced brightness in the 131~{\AA} images, so that S1 also samples the expansion of the elbow in this moderate-velocity expansion phase. The speed sharply increases to 100--150~km\,s$^{-1}$ after $\approx$01:55~UT, which can be measured in these plots only until a few minutes after 02~UT. Assuming a radial ascent, the projected speed of the sigmoid's middle part along slit S1 translates to a radial speed of $\approx$410~km\,s$^{-1}$. Unfortunately, the sigmoid fades too quickly to reveal the accelerated nature of its motion in these plots. However, by analogy with previous observations of hot channels and due to the absence of any other moving low coronal structure that can be associated with the launch of the CME, it appears certain that the fast rise of the hot channel after 01:55~UT represents the launch of the CME, implying a further acceleration up to the observed CME speed. 

This is corroborated by the combined timing and position data of the hot channel and CME front and by their comparison with the X-ray light curves, as shown in Figure~\ref{f:lightcurves}. The combined position data can be well fit by a single function implying acceleration, for simplicity here chosen to be a quadratic function, which yields an average acceleration of order $\sim$320~m\,s$^{-2}$. Most of the CME acceleration occurs during the rise phase of the long-duration flare, up to the first two LASCO data points. The LASCO data alone can be well fit by a linear function (green dashed line), which is superior in the CME propagation phase at greater heights and yields a CME speed of $\approx$1300~km\,s$^{-1}$. The timing of the acceleration inferred from the fits of the position data is very well in line with the typical behavior that the main acceleration of CMEs is synchronized with the impulsive rise of the soft X-ray light curve of their associated flare \citep{ZhangJ&al2001, Neupert&al2001, Maricic&al2007, Temmer&al2008, Temmer&al2010, ChengX&al2020}. The long duration of the eruptive flare's impulsive phase is consistent with the large spatial scale of the erupting flux indicated by the hot channel, i.e., with the fact that one foot point lies in the remote unipolar spot.

It is important to note that magnetic reconnection in the vertical current sheet spawned by the final precursor and first impulsive flare continues throughout the compound event, i.e., also between the impulsive and long-duration flares. This follows from the continuous feeding of the main flare ribbons within the bipolar spot group, the continuous presence of the nonthermal 17~GHz emission source between them (Figure~\ref{f:ribbons}), the 9~GHz light curve, and from the enhanced level of the 6--12 and 12--25~keV hard X-rays at the end of the \textsl{RHESSI} spacecraft night, which is near the onset time of the long-duration flare (Figure~\ref{f:lightcurves}). The reconnection also continues to feed the flux of the rope at the expense of the overlying flux, driving it to the point it can no longer be confined by the overlying flux and a CME and associated flare set in. Such ``flux imbalance'' \citep{Bobra&al2008, SuYN&al2011} was shown to be equivalent to the torus instability \citep{Savcheva&al2012, Kliem&al2013}. This ideal MHD instability drives fast reconnection \citep{Torok&Kliem2005, Kliem&Torok2006}, which explains the associated flare. No other process leading to an amplification of reconnection around 01:55~UT can be found in the data. Amplification by itself is extremely unlikely, because the soft X-ray flux and the 9~GHz microwave flux, which indicate the level of reconnection, show a monotonically decreasing trend after the peak of the second impulsive flare, apart from very minor fluctuations of the microwave flux. This trend also rules out that the stationarity of the hot channel's central part is a projection effect: amplifying reconnection would be expected if the entire hot channel were rising already in this phase.

\section{Magnetic Field Structure and Interpretation of the Event}\label{s:magnetic}

%~~~~~~~~~~~~~~~~~~~~~~~~~~~~~~~~~~~~~~~~~~~~~~~~~~~~~~~~~~~~~~~~~~~~~~~~~~~~~~~
\begin{figure}[t!]                                                         % f06
 \centering 
 \includegraphics[width=\linewidth]{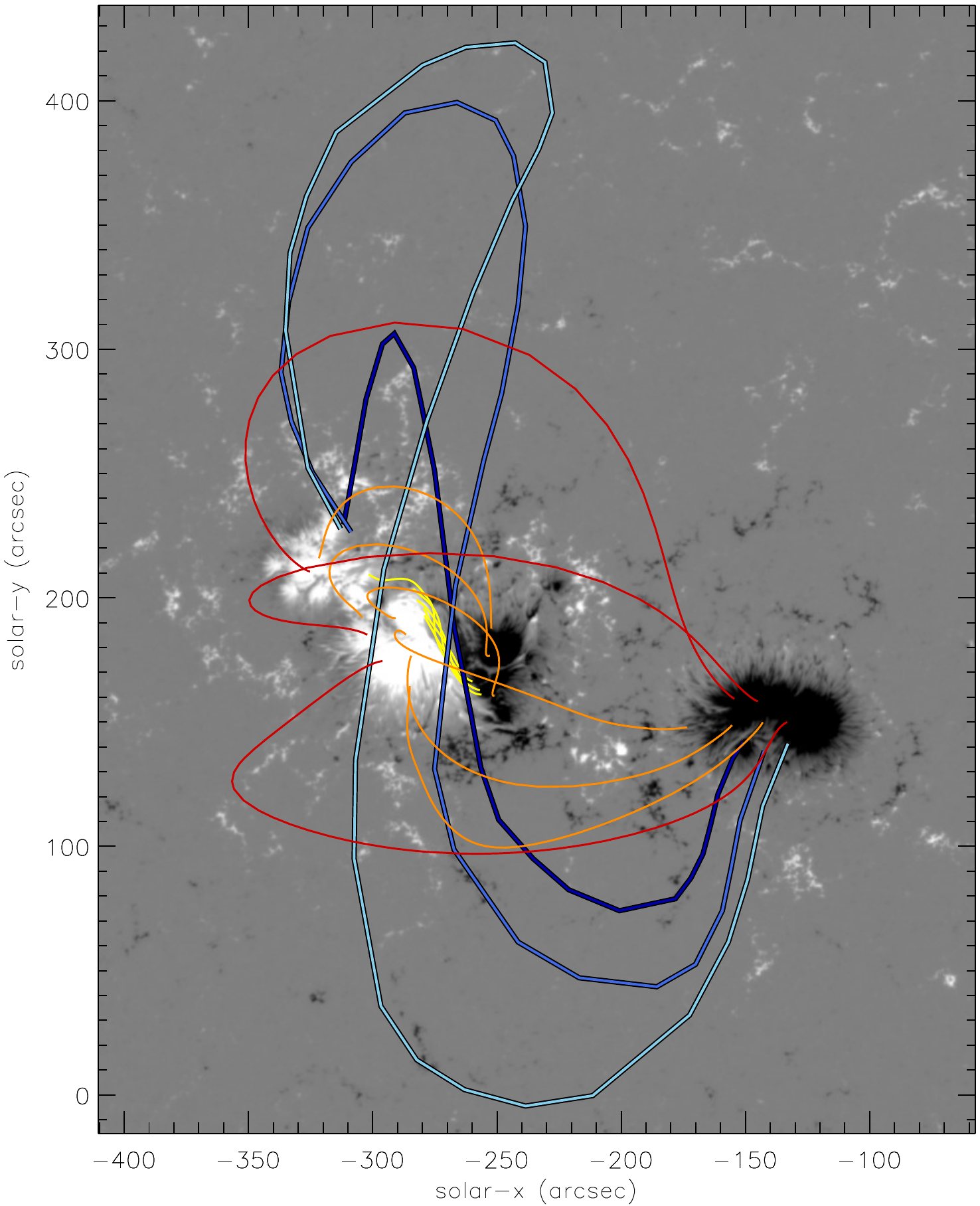}  %<-- 2-column style
 % \includegraphics[width=.75\linewidth]{mgm+fl+HC_1000Gauss.eps}
 % % \includegraphics[width=\linewidth]{flines_ns.eps} %onecolumn layout
 % % \includegraphics[width=  \linewidth]{2015jun21T0058_flb_FRyellow+amb_az00_ax90.png} 
 % % \includegraphics[width=.8\linewidth]{2015jun21T0058_flb_FR+amb_az60_ax10.png} 
\caption{HMI line-of-sight magnetogram of Active Region 12371 at 00:58:25~UT, saturated at $\pm$1000~Gauss. The evolving trace of the second (main) hot channel from Figure~\ref{f:sigmoid} (blue colored lines) and representative field lines of the NLFFF model are overlaid. The field lines show highly sheared and moderately twisted core flux along the PIL within the bipolar spot group (yellow; also shown in Figure~\ref{f:BP}) and overlying flux, orange at moderate and red at large heights.} 
\label{f:flines}
\end{figure}
%~~~~~~~~~~~~~~~~~~~~~~~~~~~~~~~~~~~~~~~~~~~~~~~~~~~~~~~~~~~~~~~~~~~~~~~~~~~~~~~

A model of the coronal magnetic structure of Active Region 12371 just prior to the event is computed utilizing the optimization code for nonlinear force-free field (NLFFF) extrapolation by \citet{Wiegelmann2004} and \citet{Wiegelmann&al2012} and the vector magnetogram from the Helioseismic and Magnetic Imager \cite[HMI,][]{Scherrer&al2012} on board \textsl{SDO} as the boundary condition. Specifically, the Spaceweather HMI Active Region Patch (SHARP) image at 00:58:25~UT is used. Standard preprocessing of the magnetogram \citep{Wiegelmann&al2006} is applied, leading to only minor changes in the present case. Three groups of selected field lines are plotted in Figure~\ref{f:flines} as yellow, orange, and red lines. These show highly sheared and moderately twisted core flux along the PIL of the bipolar spot group (yellow; $z\!<\!5$~Mm), overlying flux at moderate heights, which displays moderate shear (orange; $z\!<\!40$~Mm), and high overlying field lines, which are nearly potential (red; $z\!<\!120$~Mm). The shapes of the second (main) hot channel are overlaid: the three lines tracing the channel in Figure~\ref{f:sigmoid}(b)--(d) are copied here as blue lines with varying color depth. 

The NLFFF structure is consistent with the above interpretation that a flux rope forms or is enhanced by reconnection under the flux erupted in the final precursor and first impulsive flare, and suggests a clear picture of the event's overall evolution, driven by magnetic reconnection and flux rope instability. 
(1) The highly sheared core flux (yellow lines) is obviously the source of the precursor and first impulsive flare, as this flux runs between the prominent flare ribbons and has nearly the same extension along the PIL as the 131~{\AA} core source. This flux experiences a partial eruption, with the upper, erupting part showing up as the first hot channel and the lower, remaining part holding the core filament. 
(2) The flux at moderate heights (orange lines) does not erupt during the first impulsive flare, so that it acts as overlying flux up to this phase, confining the eruption of the core flux. 
(3) Part of the flux at moderate heights is rooted near the core flux, so that it is likely dragged into the vertical current sheet spawned by the rising core flux and reconnects there. The reconnected flux in the upward reconnection outflow wraps around the erupted core flux, forming or enhancing the flux rope which is observed as the second hot channel. One can see that the initial position and extent of the second hot channel is quite similar to the orange field lines, in particular, most of their remote foot points lie near the ends of this hot channel. 
(4) The reconnected part of the flux shown by the orange field lines, now part of the flux rope, immediately starts a dynamic evolution, seen as a strong expansion of the elbows and associated with the transient extension of the flare ribbons during the second impulsive flare. (5) The flux rope erupts fully only during the long-duration flare. This is indicated by the eruption of the entire second hot channel and consistent with the strong and persistent extension of the ribbons into the whole area under the orange field lines throughout the rise and peak phases of the long-duration flare (Figure~\ref{f:ribbons}(f)). 
Unfortunately, the first hot channel is no longer discernible by this time, so that its implied disruption cannot be verified in the data. 

\begin{figure}[t!]                                                         % f07
\centering 
\includegraphics[width=\linewidth]{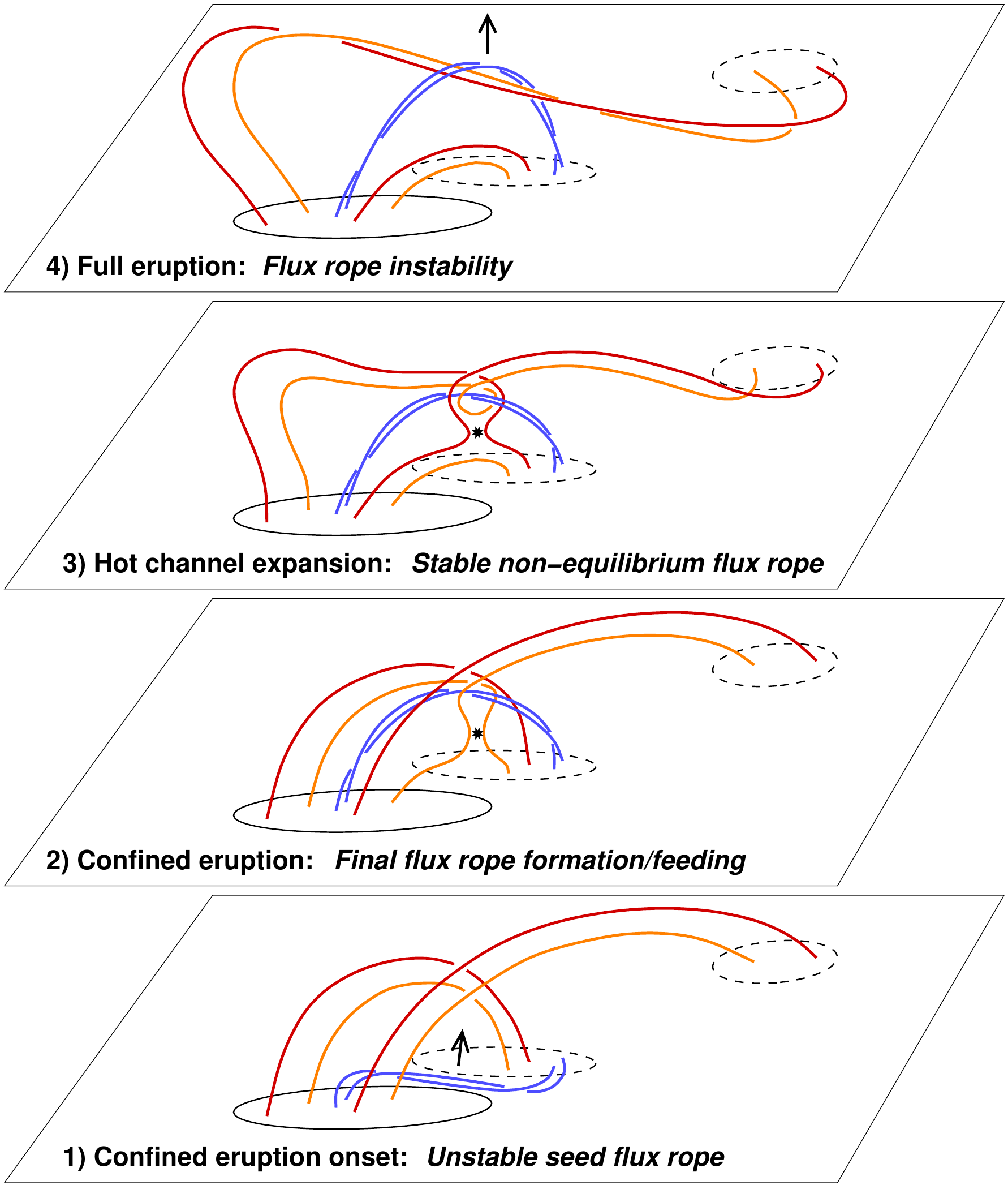}  %<-- 2-column style
\caption{Schematic of flux rope enhancement or formation by a confined eruption, in the specific geometry of the considered event. 
 (1) Unstable flux, most likely a seed flux rope, erupts but is halted by sheared overlying flux. 
 (2) ``Flare'' reconnection (asterisk) of the confined eruption, topologically identical to ``tether-cutting'' reconnection, transforms overlying flux, enhancing the erupted flux rope or forming a new one. 
 (3) This dynamically formed flux rope is in a non-equilibrium state, thus, expanding in seeking a new equilibrium, but still remains in the stable height range of the torus instability. 
 (4) Ongoing reconnection, driven by the non-equilibrium flux rope, further enhances the flux rope which eventually reaches the torus-unstable height range and then drives a CME.} 
\label{f:schematic}
\end{figure}

Additionally, the southern group of orange field lines is nearly cospatial with a set of loops at coronal temperature, best visible in the AIA bands sensitive to $\sim$1--2~MK, see Figure~\ref{f:aia304+193}(c). The southern elbow of the hot channel forms nearly cospatial with these loops, and then both expand jointly during the second impulsive flare (Figure~\ref{f:aia304+193}(d)). This additionally suggests that the orange field lines represent the coronal field realistically and join the flux rope which becomes visible as the main hot channel. Unfortunately, these loops fade even earlier than the hot channel (during $\sim$01:50--02:05~UT, progressively from lower to higher temperature), so that one cannot see their conjectured eruption in the course of the long-duration flare. However, the fading ordered by temperature suggests heating as its cause (in addition to expansion), which supports the picture that part of the loops which are cospatial with the orange field lines reconnect during the second impulsive flare. The suggested sequence of eruptions and reconnection is summarized in the schematic of Figure~\ref{f:schematic}. 

%~~~~~~~~~~~~~~~~~~~~~~~~~~~~~~~~~~~~~~~~~~~~~~~~~~~~~~~~~~~~~~~~~~~~~~~~~~~~~~~
%~~~~~~~~~~~~~~~~~~~~~~~~~~~~~~~~~~~~~~~~~~~~~~~~~~~~~~~~~~~~~~~~~~~~~~~~~~~~~~~

\subsection{Bald Patches}\label{ss:BP}

\begin{figure}[t!]                                                         % f08
\centering 
\includegraphics[width=\linewidth]{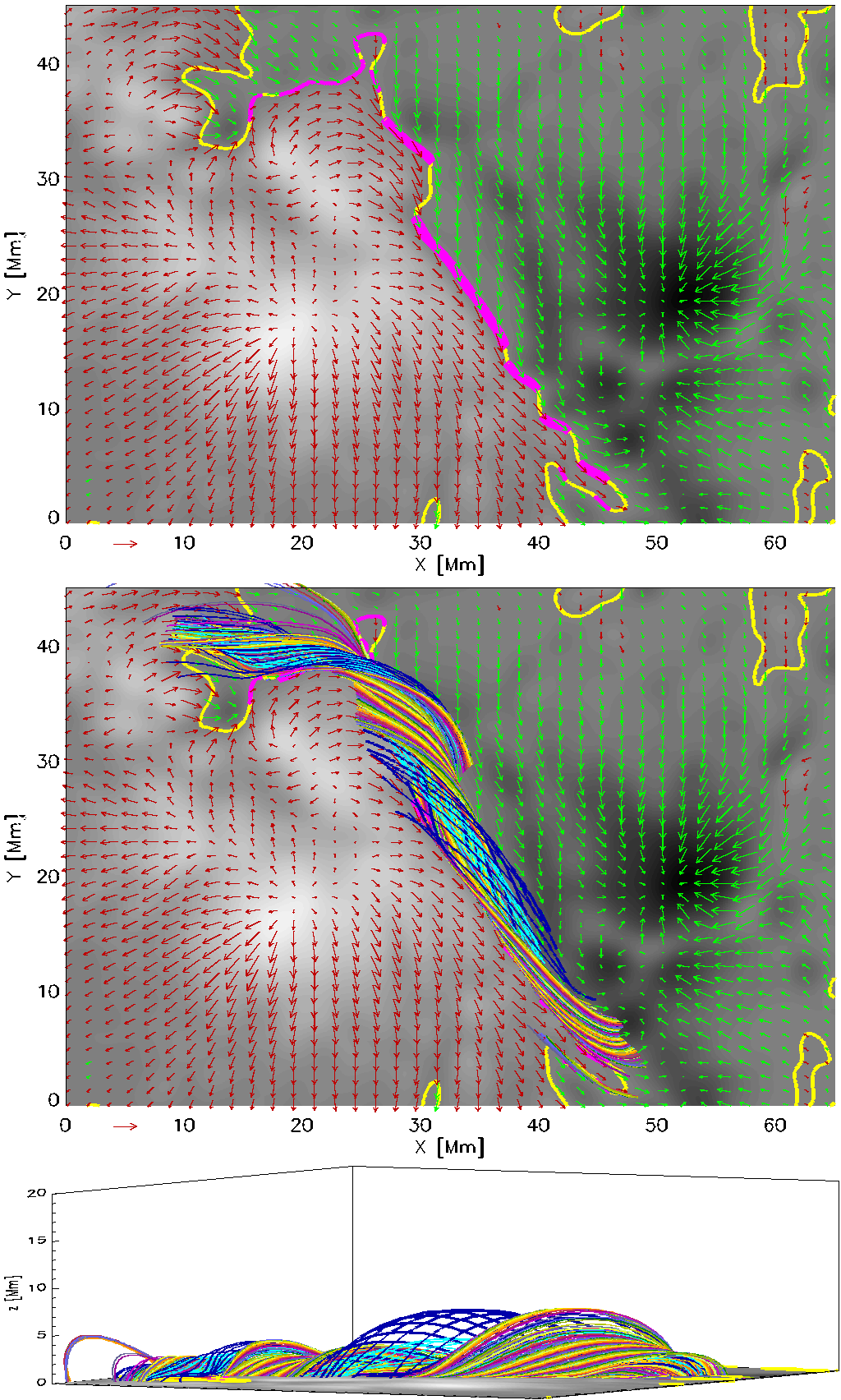}  %<-- 2-column style
\caption{
 (a) Preprocessed HMI SHARP\_CEA image of the event's 
core region at 00:58:25~UT, showing $B_z$ in gray scale, saturated at 1500~Gauss, horizontal field vectors in red and green, the PIL in yellow, and its bald-patch sections in magenta. The arrow in the bottom margin of the plot is scaled to a horizontal field strength of 1500~Gauss. 
 (b--c) Rainbow-colored field lines started in the bald patches at the main section of the PIL ($4<y<35$, marked with the thicker line in (a)) lie in the surface of a forming flux rope; field lines started in the two bald patches in the strongly bent section of the PIL further southeastward outline two arcades that meet at the bald patches. The two main sections of the forming flux rope are visualized through cyan-colored field lines started at vertical ellipses of $\approx$3/4 the respective radii near the section's apex points ($\mathbf{x}=(35,21,2)$ for the large section and $\mathbf{x}=(20.5,39,1.3)$ for the small section) and through dark blue-colored field lines near the surface of the rope sections.} 
\label{f:BP}
\end{figure}

To address the structure of the active region's core field, we first check the magnetogram for the existence of bald patches. We analyze HMI SHARP data taken during 00:46--07:22~UT (the end of the undisturbed decay of the long-duration flare), determining the bald-patch sections of the PIL as defined in \citet{Titov&al1993}, and find very similar results throughout this period. Using the preprocessed magnetogram at 00:58:25~UT, Figure~\ref{f:BP}(a) shows that the bald-patch topology exists along most of the relevant ``main'' section of the PIL ($27<x<46$, $4<y<35$), where a filament channel indicates energy storage and where the first hot channel, the microwave source, and the first ribbons and flare loops form. Nearly identical locations of the PIL and bald patches are found in the original HMI SHARP magnetogram. This provides a strong indication that a transition from arcade to flux rope structure has already significantly progressed at this section of the PIL prior to the first impulsive flare. The field lines passing here in the inverse direction must cross the PIL again at either end because there is not a sufficient amount of opposite-polarity flux at the same side of the PIL. The situation is different for the bald patches in the adjacent strongly curved sections of the PIL further southeastward ($y<5$), which partly also possess a bald-patch topology. Here two alternating pairs of opposite polarity exist in relatively close proximity, embedding three relatively closely spaced PIL sections, so that the bald patches in the middle PIL section can be the boundary between two field line arcades that pass over the outer PIL sections \citep[as sketched in Figures~1(a) and 2 in][]{Titov&al1993}. We emphasize that the existence of a forming flux rope above the relevant section of the PIL is directly indicated by the structure of the vector magnetogram. 

The computed NLFFF model, shown in Figure~\ref{f:BP}(b)--(c), exactly conforms to this conclusion and further suggests that a relatively coherent flux rope exists already above most of the relevant PIL section, i.e., at $4\lesssim y\lesssim35$, only the short interval at $27<y<32$ is still of arcade structure. The field lines in the core of the forming flux rope (colored cyan in Figure~\ref{f:BP}) show a twist of $N\approx0.5\mbox{--}1$ turns. A second, smaller section of a forming flux rope is indicated at $y\sim40$, above bald patches in the adjacent part of the PIL. Only a minor part of the flux in these two sections of the forming flux rope extends along the whole length of the structure, as can be seen in Figure~\ref{f:flines} which shows only the large section (there colored in yellow). The flux rope was also found by \citet{Mitra&al2020}, with minor differences, e.g., apparently less flux content, in their NLFFF computation using the same code. 

The existence of a forming flux rope suggests that the ideal MHD model may also be relevant for the first impulsive flare and its precursors. Moreover, the presence of bald patches under most of the core filament shown in Figure~\ref{f:aia304+193}(a) is consistent with the fact that the filament does not erupt. As numerically demonstrated by \citet{Gibson&Fan2006} and observationally supported by \citet{ChengX&al2018}, the photospheric line-tying in a bald patch prevents the lower part of a flux rope from participating in eruptions; only the upper part tears off and erupts when reaching the condition for instability. Thus, if the filament is trapped sufficiently low in the flux above the bald patches, it can survive an eruption.  

%~~~~~~~~~~~~~~~~~~~~~~~~~~~~~~~~~~~~~~~~~~~~~~~~~~~~~~~~~~~~~~~~~~~~~~~~~~~~~~~
%~~~~~~~~~~~~~~~~~~~~~~~~~~~~~~~~~~~~~~~~~~~~~~~~~~~~~~~~~~~~~~~~~~~~~~~~~~~~~~~

\subsection{Field Decay Index, Confinement, and Eruption}\label{ss:decayindex}

Since the flux erupting in the first impulsive flare (indicated by the first hot channel in Figure~\ref{f:sigmoid}(a)) does not show any writhing due to the helical kink instability, an interpretation of the event in terms of flux rope instability requires that both the initial confined and eventual ejective eruptions, as well as the intermediate confinement, be consistent with the properties of the torus instability \citep{Kliem&Torok2006}. Stability and confinement are expected if the decay index of the external field at the position of the flux rope axis is subcritical, $n=-d\ln{B_\mathrm{ep}}/d\ln{z}<n_\mathrm{cr}$, and a full (ejective) eruption is expected for supercritical decay index, $n>n_\mathrm{cr}$. Here the poloidal component of the external field, $B_\mathrm{ep}$, is relevant, which can be approximated by the poloidal component of the potential field in the plane perpendicular to the flux rope's direction of propagation, $B_{\mathrm{pp}}$. This is more appropriate than the approximation by the whole perpendicular component of the potential field, which is often used. Assuming that reconnection can readily proceed under the flux rope, which is suggested for the present event by the prompt formation of flare ribbons and loops, the critical decay index lies around the canonical value of $n_\mathrm{cr}\sim3/2$ (\citealt{Kliem&Torok2006}; also see the discussion in \citealt{Kliem&al2013} and \citealt{Zuccarello&al2015, Zuccarello&al2016}). We adopt the canonical value for the purpose of the present semi-quantitative comparison with the available data, which lack the knowledge of the height ranges for the confined and ejective eruptions. The potential field is computed from the original HMI SHARP\_CEA image using the Green function \citep{Schmidt1964}. The resulting decay index profiles are indistinguishable from the ones computed with the preprocessed magnetogram.

%~~~~~~~~~~~~~~~~~~~~~~~~~~~~~~~~~~~~~~~~~~~~~~~~~~~~~~~~~~~~~~~~~~~~~~~~~~~~~~~
\begin{figure}[t!]                                                         % f09
\centering 
\includegraphics[width=\linewidth]{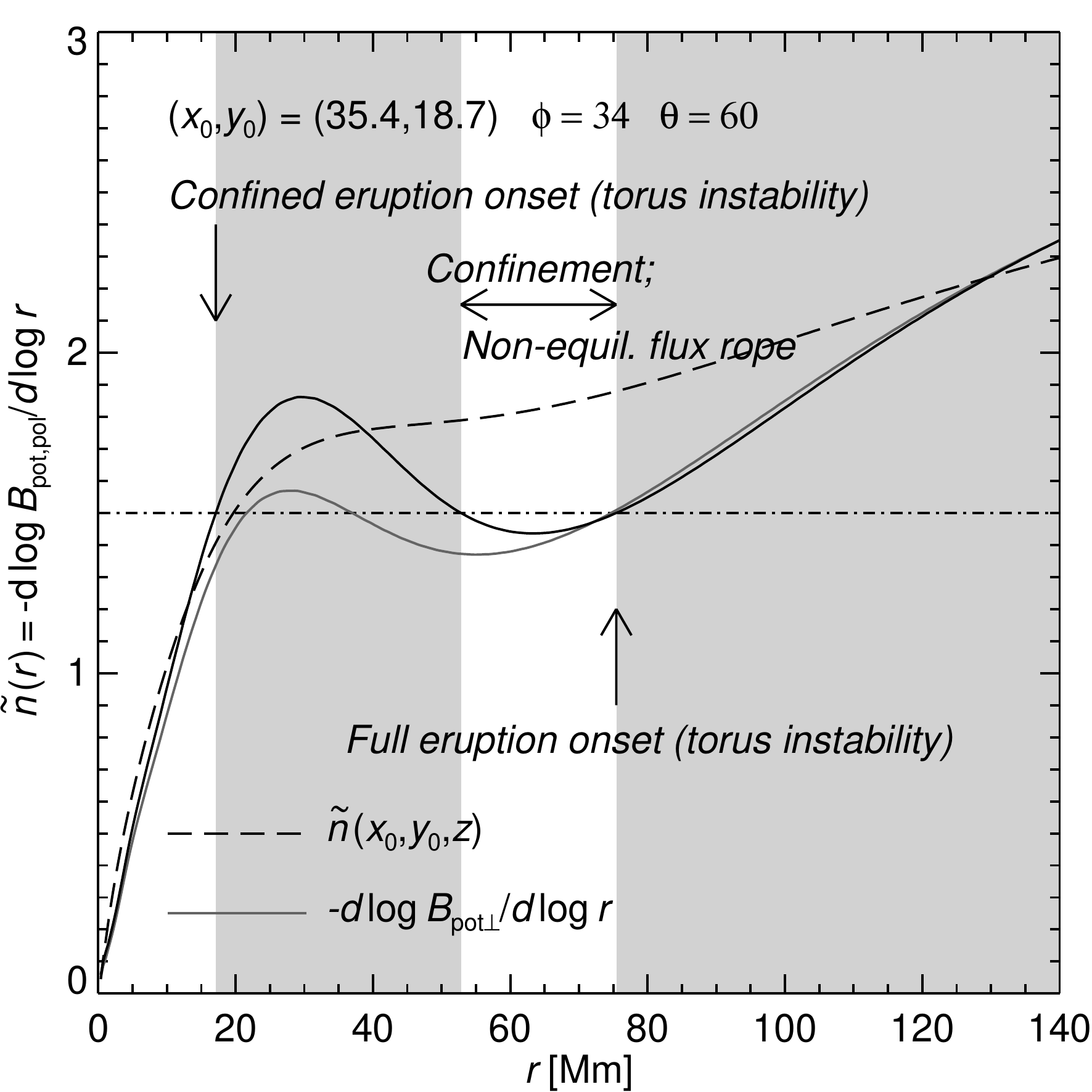}  %<-- 2-column style
\caption{Oblique ($\tilde{n}(r)$) and vertical ($\tilde{n}(z)$) decay index profile of Active Region~12371 on 2015 June 21, 00:58~UT, starting at the point given by solar-$(x_0,y_0)=(-270,180)$~arcsec in Figure~\ref{f:flines} ($(35.4,18.7)$~Mm in Figure~\ref{f:BP}) and $z=0$. The oblique profile is taken along a direction $\bm{\hat{r}}$ inclined by $30^\circ$ from the vertical to the west side of the PIL in the direction perpendicular to the PIL ($\phi=34^\circ$, $\theta=60^\circ$). The poloidal component of the potential field in the plane perpendicular to the chosen direction ($\bm{\hat{r}}$ or $\bm{\hat{z}}$), $B_{\mathrm{pp}}$, is used. Torus-unstable height ranges of the $\tilde{n}(r)$ profile are shown shaded. For comparison, the decay index along the inclined direction, computed using the whole perpendicular component of the potential field, $B_{\mathrm{pot}\perp}$, is plotted in gray.} 
\label{f:decayindex}
\end{figure}
%~~~~~~~~~~~~~~~~~~~~~~~~~~~~~~~~~~~~~~~~~~~~~~~~~~~~~~~~~~~~~~~~~~~~~~~~~~~~~~~

Figure~\ref{f:decayindex} shows the resulting standard vertical decay index profile, $\tilde{n}(z)=-d\ln{B_{\mathrm{pp}}}/d\ln{z}$, and the profile along an oblique propagation direction, $\bm{\hat{r}}$, inclined to the west side of the PIL in the direction perpendicular to the PIL, i.e., $\tilde{n}(r)=-d\ln{B_{\mathrm{pp}}}/d\ln{r}$. This is the natural direction for an inclined rise, and only this direction allows a natural choice for the toroidal direction in the perpendicular plane (parallel to the PIL) and, hence, the poloidal direction (perpendicular to the toroidal direction) in this plane. A westward-inclined rise of the unstable flux is indicated by the westward displacement of the first hot channel (Figure~\ref{f:sigmoid}(a)). This is consistent with the fact that the immediate environment of the relevant section of the PIL (Figure~\ref{f:BP}) shows a larger amount of (positive) flux on the east side compared to the (negative) flux on the west side; such imbalance is known to deflect erupting flux toward the weak-field side \citep{Panasenco&al2013}. The inclination must be quite substantial ($>\!20^\circ$ from the vertical) because the eruption originates $\approx\!17^\circ$ east of central meridian. One can see that the decay index profile $\tilde{n}(r)$ indeed possesses the structure that allows both confined and ejective eruptions, i.e., an inflection point introducing a finite stable height range above an unstable one, located between the standard ranges of stability at small heights and instability at large heights, where the decay index approaches its asymptotic value of 3. The decay index profiles computed from the whole perpendicular component of the potential field in the plane perpendicular to the direction of propagation, $B_{\mathrm{pot}\perp}$, show the same structure, but the lower unstable range in the $\tilde{n}(r)$ profile is narrower and less supercritical. We find this structure for inclination angles of $(90-\theta)\gtrsim20^\circ$ from the vertical and in a broad range of position angles of $-60^\circ\lesssim\phi\lesssim45^\circ$ around the westward direction. 

The toroidal component of the potential field at the lower critical distance along the inclined direction $\bm{\hat{r}}$, $r_\mathrm{cr1}=17.1$, amounts to 0.79 of the poloidal component. A toroidal component acts stabilizing \citep{Kliem&al2014b}, but here its magnitude is not sufficient to remove the lower torus-unstable range, as indicated by the decay index profile computed from $B_{\mathrm{pot}\perp}$, which has the toroidal field component folded in. 

Such a decay index profile with two unstable height ranges, first pointed out in \citet{GuoY&al2010a} and found to be not uncommon in multipolar active regions \citep{WangD&al2017}, results naturally in multi-polar source regions \citep{Torok&Kliem2007}, but also from the presence of two sufficiently different spatial scales in the active-region structure for simple bipolar regions, as in Active Region~12371. (This has been pointed out independently by \citealt{Filippov2018}.) In a bipole, the decay index reaches 3/2 at the height $z=L$, where $L$ is the half-distance between the poles. If the main flux concentrations in the magnetogram are arranged on two clearly separate scales $L_1$ and $L_2$, then the decay index profile can show a superposition of two slopes (both starting near the origin) which reach $n(z)=3/2$ at, roughly, $z\sim L_1$ and $z\sim L_2$. Active Region~12371 displays such a two-scale structure, with the smaller scale, $L_1\sim 15$~Mm, given by the distance from the PIL of the two bigger sunspots closest to the PIL (Figure~\ref{f:BP}) and the larger scale, $L_2\sim85$~Mm, given by the distance of the separate unipolar spot (Figure~\ref{f:flines}). These values correspond quite well to the distances from the PIL of $r_\mathrm{cr1}=\bm{17}$~Mm and $r_\mathrm{cr2}=\bm{75}$~Mm along the observationally indicated direction $\bm{\hat{r}}$ in Figure~\ref{f:decayindex} where $\tilde{n}(r)=3/2$. A close agreement, i.e., $r_\mathrm{cr1,2}\approx L_{1,2}$, cannot be expected because each of the values $r_\mathrm{cr1,2}$ depends on the whole flux distribution in the magnetogram. The vertical direction does not run through the center of the outer bipole in the active region, so that the influence of the large scale is only weakly indicated in the decay index profile $\tilde{n}(z)$. 

The decay index profile $\tilde{n}(r)$ in Figure~\ref{f:decayindex} allows for confined eruptions from $\lesssim$20~Mm (ending in the range $\sim$50--75~Mm) and for ejective eruptions (CMEs) from $\sim$75~Mm. This start height for confined eruptions lies considerably above the apex height of the forming flux rope of $\sim$5~Mm in the NLFFF coronal field model (Figure~\ref{f:BP}). However, one has to keep in mind that the position and flux content of the flux rope in the NLFFF is not very precise. For most NLFFF extrapolation codes, including the one used here, it is generally difficult to obtain a high-arching flux rope. 

For an inclination of $30^\circ$ from the vertical, the projected distance of $\sim$25~arcsec between the upper edge of the first hot channel and the PIL at E17 (Figure~\ref{f:sigmoid}(b)) corresponds to a true distance $r\sim82$~Mm. This is consistent with the upper edge of the upper stable range of the $\tilde{n}(r)$ profile in Figure~\ref{f:decayindex}. 

The decay index profile $\tilde{n}(r)$ also suggests that mostly the flux visualized by the northern set of orange field lines in Figure~\ref{f:flines} must confine the initially erupting flux. Their apex points have distances of $\approx$25--40~Mm from the PIL along the inclined direction used in Figure~\ref{f:decayindex}, which places them below the upper stable height range, so that they can contribute to forming or enhancing a stable flux rope in this range when stretched upward by the confined eruptions. Therefore, the shear, orientation, and height range of the orange field lines in Figure~\ref{f:flines}, combined with the corresponding oblique decay index profile in Figure~\ref{f:decayindex}, support our interpretation of the event in terms of flux rope enhancement by the flare reconnection of the confined eruptions, followed by a CME/eruptive flare due to flux rope instability. 

In the upper unstable height range, the flux distribution of the active region is much less asymmetric, so that the ejective eruption can proceed in approximately vertical direction, as indicated by the eastward displacement of the second hot channel (Figure~\ref{f:sigmoid}(d)).

\subsection{Post-eruption Magnetic Structure}\label{ss:post_eruption}

%~~~~~~~~~~~~~~~~~~~~~~~~~~~~~~~~~~~~~~~~~~~~~~~~~~~~~~~~~~~~~~~~~~~~~~~~~~~~~~~
\begin{figure}[t!]                                                         % f10
\centering 
\includegraphics[width=\linewidth]{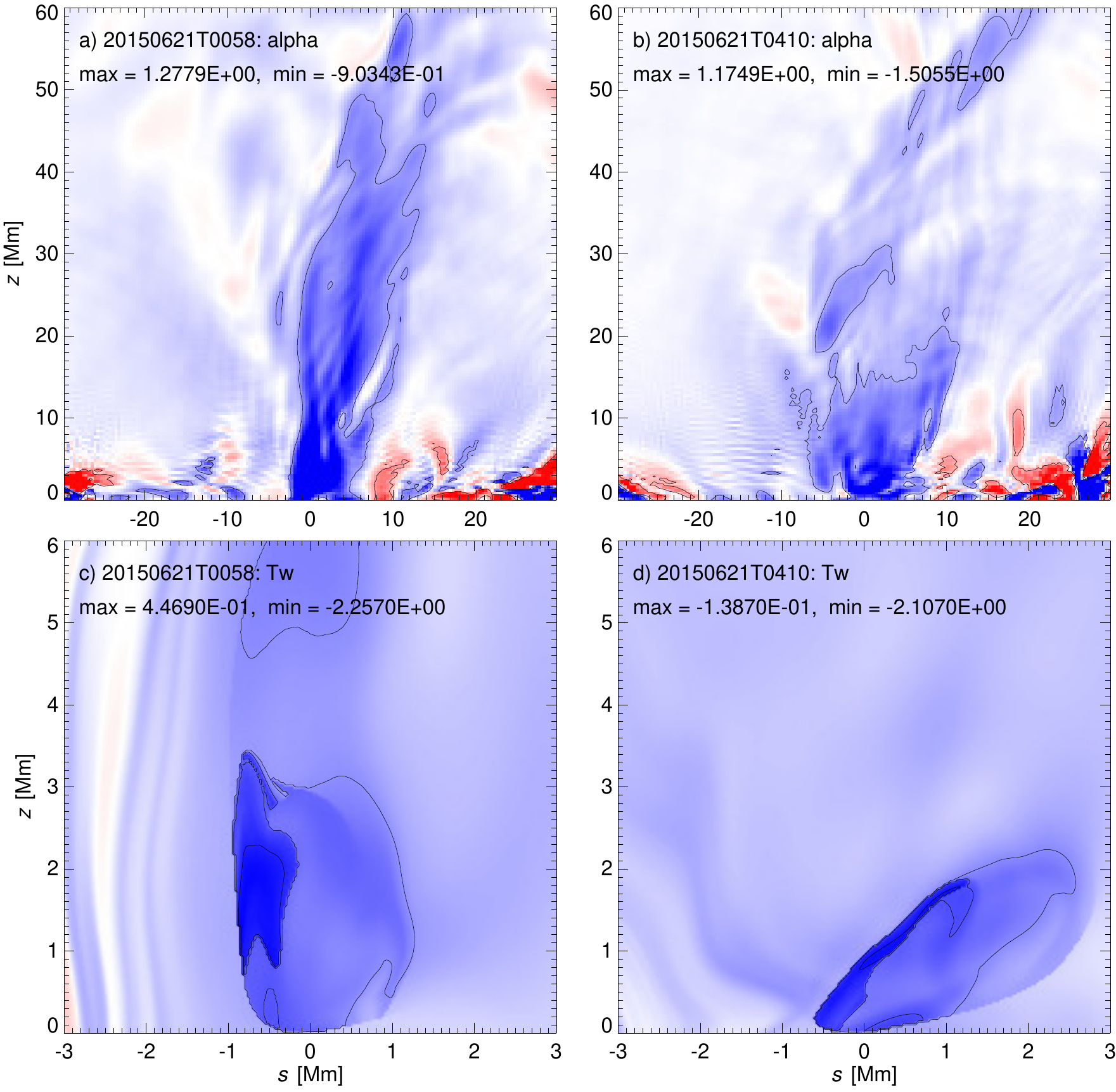}  %<-- 2-column style
\caption{Vertical cut showing the force-free parameter $\alpha$ in the center of the active region at (a) 00:58~UT and (b) 04:10~UT. The point $s=0$ corresponds to $(x,y)=(35.4,18.7)$~Mm in Figure~\ref{f:BP}, and the cut plane is rotated by $60^\circ$ from the $x$ axis for optimum display of the vertical current layer. Contours are drawn at $\alpha=\pm0.1$~Mm$^{-1}$. (c)--(d) Same for the twist parameter $T_w$, displaying the cross section of the flux rope in the core of the active region (with an optimum rotation angle of the cut plane of $20^\circ$). The contours $T_w=-1$, $-1.5$, and $-2$ are drawn.} 
\label{f:NLFFF}
\end{figure}
%~~~~~~~~~~~~~~~~~~~~~~~~~~~~~~~~~~~~~~~~~~~~~~~~~~~~~~~~~~~~~~~~~~~~~~~~~~~~~~~

We have considered the HMI magnetograms also at 01:58, 02:58, 04:10, and 07:22~UT and computed an NLFFF model for the latter two times. These are taken from the decay phase of the eruption, so that the NLFFF assumption should again be relatively well justified, and at the same orbital position of \textsl{SDO} as the magnetogram at 00:58~UT, minimizing the orbital artifacts \citep{Schuck&al2016}. The 1--8~{\AA} soft X-ray flux decreases to 15~percent of the peak value by 04:10~UT and to 10~percent by 07:22~UT, but a further confined eruption associated with a C2 flare occurs in the region close to Ribbon R$_3$ at 07:03~UT, and the NLFFF at 07:22~UT already shows signs of new energy storage.

The magnetograms, especially the bald-patch sections of the main PIL in the active region core, show little change in this whole interval. Correspondingly, the NLFFF models are rather similar, too. In particular, the field line plots are very similar to Figure~\ref{f:flines} if the appropriate (slightly shifted) start points are chosen for the core flux and identical start points are chosen for the two sets of overlying field lines. This suggests the reformation of the core flux rope after the main eruption, as expected from its bald-patch topology \citep{Gibson&Fan2006}. Figure~\ref{f:NLFFF}(c)--(d) and Table~\ref{t:t1} show that the flux rope at 04:10~UT has a smaller flux content and height than the original flux rope, which suggests that the original flux rope was strongly perturbed during the eruption, i.e., most likely erupted partially. The reformed rope reaches a similar twist. The twist is here measured by the twist parameter $T_w$ defined in \citet{Berger&Prior2006}, which reaches up to $|T_w|\sim2$, higher than the number of field line turns suggested by the field line plot in Figure~\ref{f:BP}, $N\sim0.5\mbox{--}1$. This difference is similar to a previous case and does not present a contradiction \citep{LiuR&al2016}. By 07:22~UT, the reformed flux rope reaches the same flux content and height range as the original rope at 00:58~UT (not shown), suggesting new energy storage. 

An interesting aspect is the presence of a diffuse vertical current layer in the sheared flux above the core of the active region, which indicates the changes due to the eruption more clearly than a field line plot. The layer has a width of $\sim$10~Mm, a height of $\sim$60~Mm, is westward inclined, and possesses irregular substructure (Figure~\ref{f:NLFFF}(a)--(b)). At 04:10~UT (Figure~\ref{f:NLFFF}(b)), the current layer is strongly perturbed by the eruption and shows a more irregular distribution of the current density, which is also more concentrated at lower heights, equivalent to a reduction of the contained free magnetic energy. It is noteworthy that the perturbation is strongest in the lower torus-unstable height range and that it extends down to the core flux rope. 

The large-scale properties of the NLFFF, compiled in Table~\ref{t:t1}, suggest that the free magnetic energy decreases by $2.5\times10^{31}$~erg during the compound event, consistent with its magnitude \cite[a relatively fast CME and low M-class flares;][]{Emslie&al2012}. The total energy and relative magnetic helicity also decrease. Flux cancellation continues throughout the event, leading to a decrease of the potential-field energy. By 07:22~UT, the photospheric magnetic flux and potential-field energy have decreased further, but the total and free energy and relative helicity have begun to increase, consistent with the occurrence of the confined eruption and C2 flare near this time. The axial flux of the core flux rope shows a similar evolution: 36~percent are lost by 04:10~UT but restored by 07:22~UT. 
% \newline 

%*******************************************************************************
\begin{table}[t]                                                        % Tab. 1
\caption{Large-scale properties of the NLFFF model before and after the eruption. $W_\mathrm{NLFFF}$ is the magnetic energy of the NLFFF, $W_\mathrm{pot}$ the corresponding potential-field energy, $W_\mathrm{virial}$ the energy of the NLFFF estimated from the virial theorem, and $W_\mathrm{free}=W_\mathrm{NLFFF}-W_\mathrm{pot}$. $F_\mathrm{u}$ is the unsigned flux in the magnetogram area (slightly smaller than the respective HMI SHARP\_CEA images), $F_\mathrm{FR}$ the total axial flux of the flux rope in the core of the active region within the $T_w=-1$ contour, computed in the cross section shown in Figure~\ref{f:NLFFF}(c)--(d), and $H_\mathrm{m}$ is the relative magnetic helicity of the NLFFF, obtained using the expression in \citet{Valori&al2012}.}
\vspace{12pt} 
\begin{tabular}{lccc}
\tableline 
Time                          & 00:58:25~UT  & 04:10:25~UT  \\ 
\tableline 
$W_\mathrm{NLFFF}$            & 2.88e+33 erg & 2.82e+33 erg \\ 
$W_\mathrm{pot}$              & 2.44e+33 erg & 2.40e+33 erg \\ 
$W_\mathrm{free}$             & 4.39e+32 erg & 4.14e+32 erg \\ 
% $W_\mathrm{free}$           & 18.0\%       & 17.2\%       \\ 
$W_\mathrm{virial}$           & 2.55e+33 erg & 2.26e+33 erg \\ 
$F_\mathrm{u}$                & 2.83e+22 Mx  & 2.78e+22 Mx  \\ 
$F_\mathrm{FR}$               & 6.56e+19 Mx  & 4.19e+19 Mx  \\ 
$H_\mathrm{m}/F_\mathrm{u}^2$ & -0.0556      & -0.0528      \\ 
\tableline 
\end{tabular}
% \tablecomments{...}
\label{t:t1}
\end{table}
%*******************************************************************************

\subsection{Flux Cancellation vs. Flux Emergence} 

Shearing and converging flows dominate at the highly sheared section of the PIL in the days preceding the event. Signs of significant flux emergence exist only in other parts of the active region, and the total flux in both polarities decreases. This implies the dominance of flux cancellation over flux emergence in the evolution toward the event. The same conclusion was reached by \citet{Vemareddy2017} and \citet{Mitra&al2020} in a dedicated investigation of the active region's magnetic evolution, where more detail can be found. Therefore, the formation of the flux ropes by flux emergence can be excluded.

\subsection{Homologous Events}\label{ss:homologous} 

The present event is one in a series of four sizable homologous events from Active Region~12371 during June~18--25. The four CMEs reach projected velocities in the range 1200--1600~km\,s$^{-1}$, and all four associated long-duration flares are of M class. Detailed studies of these and a few weaker events from the active region were performed by \citet{Vemareddy2017}, \citet{Vemareddy&Demoulin2018}, and \citet{LiuR&al2019}; also see \citet{Mitra&al2020} and references therein for further relevant literature on the individual events. These authors concluded that the large-scale structure of the active region remains similar and that the build-up of energy is mainly due to shearing and converging motions associated with flux cancellation and tether-cutting reconnection at the same section of the PIL throughout the relevant period. Their NLFFF extrapolations indicate the existence of a flux rope in formation for events 2--4, and bald patches exist in events 2--3 (the first event occurs too far east to allow firm conclusions in this regard). The third event, SOL2015-06-22T18:23, was observed in high resolution and has found particular attention \citep{JingJ&al2016, JingJ&al2017, LiuC&al2016, WangH&al2017, Sahu&al2020}. 

From the similarity of the large-scale active-region structure \citep{Vemareddy&Demoulin2018}, a similar two-scale decay index profile can be expected to govern the dynamics of the erupting flux in all four events. Event~3 (2015-06-22) also has a soft X-ray light curve of a compound event, consisting of an impulsive and a long-duration flare which strongly overlap, and is preceded by two confined, impulsive precursor flares, which are superimposed on the rise profile of the main flare and peak at 17:27 and 17:47~UT, about 35 and 15~minutes before the impulsive phase of the main flare reaches its peak. \citet{WangH&al2017} suggested that all four activities in this event were likely triggered by the observed small-scale flux emergence near the PIL. This conforms to the model by \citet{Kusano&al2012}, but the following two aspects of the event are not yet explained in the context of this model. First, the model in its present form requires the amount of emerging flux to match the erupting flux. From \citet{Bobra&al2008}, \citet{Aulanier&al2010}, \citet{Amari&al2010}, \citet{Green&al2011}, \citet{SuYN&al2011}, and \citet{Ishiguro&Kusano2017}, the latter should be of order 6--30\,\% of the unsigned active-region flux \cite[given as $\approx\!2\times10^{22}$~Mx by][]{Mitra&al2020}, whereas only $\sim(1\mbox{--}2)\times10^{19}$~Mx were observed to emerge. Second, the model does not yet explain confined eruptions. On the other hand, the model suggested here for the eruption on 2015 June~21 is consistent also with the other three homologous eruptions from the active region, because the main elements of the model existed as well. First, a decay index profile with two unstable height ranges; second, two or more confined flares which produced a hot channel prior to each CME.

\section{Discussion}\label{s:discussion}

\subsection{Ideal MHD vs.\ Reconnection Model for the CME}

Our analysis of the compound flare, associated hot channels, and CME has demonstrated the following. 
(1) The two impulsive flares remain confined. 
(2) The second (main) sigmoidal hot channel, first completely seen during the first impulsive flare at $\approx$01:30~UT, illuminates a flux rope. 
(3) The full eruption of the hot channel, which commences at $\approx$01:55~UT, is the onset of the CME. 
(4) The long-duration flare, which commences at about the same time, is associated with the CME (i.e., an eruptive flare). 
Therefore, we conclude that the eruption of a \emph{preexisting} flux rope causes the CME and its associated long-duration flare. 

Furthermore, magnetic reconnection in the vertical current sheet continues after the second impulsive flare peak up to the onset of the eruptive flare with a clearly decreasing trend. This speaks against the hypothesis of the reconnection models that a self-amplification of reconnection, i.e., amplification without an external driver, initiates and drives the CME. We conclude that this CME and associated long-duration flare conform to the ideal MHD model only. The decay index profile of the active region shows two unstable height ranges, consistent with the initiation of both confined and ejective flux rope eruptions by onset of the torus instability.

\subsection{Beginning Flux Rope Formation Prior to the Confined Eruptions}\label{ss:formation}

Regarding the question when a flux rope forms in the first instance and how the precursors and the first impulsive flare are initiated, we have found three indications of a preexisting flux rope. First, the existence of bald-patch sections along the major part of the main PIL under the erupting core flux yields a strong indication that a large part of the core flux has transitioned from arcade to flux rope topology already prior to the precursors. It is not clear how coherent the forming flux rope already is, but the core field certainly does no longer possess the topology of a magnetic arcade. Second, the NLFFF extrapolation suggests that a flux rope exists in the initially erupting core flux. Third, the diffuse envelope of the core brightening indicates, also tentatively, that a first hot channel, outlining an erupting flux rope, has formed already during these early phases of the compound event. Overall, their mechanism cannot be inferred unambiguously, but the existing, tentative indications are in favor of the ideal MHD model.

\subsection{NLFFF Model}\label{ss:NLFFFmodel}

The NLFFF model for this event supports several aspects of our analysis, in particular, (1) the formation of a flux rope in the highly sheared core flux above the bald patches at the main PIL section, where the final precursor and the impulsive flares originate (Figure~\ref{f:BP}), and (2) the existence of moderately sheared overlying flux, which can confine these eruptions and subsequently be turned into a new or enhanced flux rope, illuminated by the sigmoidal second (main) hot channel (Figure~\ref{f:flines}). However, we emphasize that all of our conclusions regarding the genesis of the CME are derived directly from the data analysis and from the potential field; none of them requires the NLFFF model. 

The conjecture that the torus instability also launches the confined eruptions is only partly supported by the NLFFF, because the flux rope does not extend into the torus-unstable height range and has a flux content of only $\sim$\,0.2~percent the unsigned flux in the magnetogram (Table~\ref{t:t1}), far lower than the $\sim$\,6--30~percent estimated to enable a full eruption (see references in Section~\ref{ss:homologous}). On the other hand, the sheared overlying flux (Figure~\ref{f:NLFFF}(a)--(b)) does provide some further support. 

At 00:58~UT, the flux passing through the layer shows an indication of twist of up to $N\sim0.5$ turn, centered at heights $z\sim30\mbox{--}35$~Mm. This weakly twisted flux is rooted in the flux concentration at $(x,y)\approx(22,30)$~Mm in Figure~\ref{f:BP}, whose curl indicates enhanced twist. There is no significant enhancement of the current density in this height range (no clear indication of a current channel), so that this flux, if taken at face value, cannot be expected to drive any of the eruptions in the compound event. However, if the true field does possess an enhanced current density in this height range, instability would be indicated there by the decay index profile (Figure~\ref{f:decayindex}), consistent with the strongest change of the current layer (Figure~\ref{f:NLFFF}(a)--(b)). From previous experience with NLFFF models obtained with the optimization method, which turned out to be very stable in our MHD code although computed from magnetograms taken shortly prior to an eruption \citep{GuoY&al2010a, GuoY&al2013, ChengX&al2014, LiuR&al2016, XueZ&al2016}, we consider it possible that the true field at 00:58~UT was much closer to the onset of the torus instability in the lower unstable range than indicated by the NLFFF model, by containing a seed flux rope of larger height or flux content or both. 

The total magnetic flux of the vertical current layer measured inside the contour $\alpha=0.1$~Mm$^{-1}$ in Figure~\ref{f:NLFFF}(a) is $1.6\times10^{21}$~Mx. This amounts to 5.5~percent of the unsigned flux in the whole magnetogram area and 11~percent of the unsigned flux in the core area excluding the separate unipolar spot in the west, sufficient to enable 
the observed confined and full eruptions. 

Another open aspect of the NLFFF model is the amount of energy change through the event. The change in energy from the virial theorem, applied to the same preprocessed magnetograms, is $\sim\!3\times10^{32}$~erg, but the NLFFF loses only $\sim\!6\times10^{31}$~erg (Table~\ref{t:t1}).

\section{Implications for Eruption Models and Forecasts}\label{s:implications}

\subsection{A General Picture of Flux Rope Formation by Confined Eruptions}\label{ss:general}

Through the continuous visibility of hot plasma throughout the four phases of flaring activity, the present observations provide insight into a path of formation for a flux rope which erupts into a major CME and associated flare. This path involves the ``flare reconnection'' of a confined eruption and is of general relevance; it is likely to be realized in a large fraction of all eruptions. 

Generally, arcade field lines (which simply make a loop over the PIL) are transformed into flux rope field lines (which wind about an axis in the corona) by reconnection in an essentially vertical current sheet. There are two standard scenarios for this process, which differ in the driver and, accordingly, in speed, but are topologically identical. ``Tether-cutting reconnection'' is driven by the shearing and converging motions of canceling photospheric flux patches and has been suggested to occur in the photosphere or low chromosphere \citep{vanBallegooijen&Martens1989} or in the low corona \citep{Moore&al2001}. The much faster ``flare reconnection'' is driven by a fast coronal process, usually assumed to be an ideal MHD instability of a flux rope \citep{LinJ&Forbes2000}. Alternatively, a resistive MHD instability operating in an arcade was proposed \citep{Karpen&al2012}. However, since neither an initial equilibrium, nor an onset condition and a growth rate were given, it remains vague what instability this might be and there is no obvious distinction between this suggestion and the \emph{assumption} of the reconnection models that self-amplifying reconnection leads to eruption. 

In the present event, slow tether-cutting reconnection is the prime candidate for the evolution up to the precursors. The bald patches indicate that this evolution presumably includes a forming flux rope, which partly erupts in the final precursor and the first impulsive flare. However, this rope is only the seed for the rope whose eruption causes the CME and long-duration flare. The flare reconnection of the confined, impulsive flares adds flux to the erupted flux, forming the eventually erupting flux rope. We note that the same conclusion was drawn by \citet{Patsourakos&al2013} from the investigation of another hot channel observed continuously between a CME and its preceding confined eruption. Flux rope formation by reconnection in the corona has been suggested by a number of recent investigations, but several of them did not clearly establish the association with a confined flare \citep{LiuR&al2010, ZhangQH&al2014, ZhangQH&al2020, James&al2017, YanXL&al2018, GouT&al2019}. Even more recently (while this paper was being revised), several authors pointed out the role of confined flares in forming or significantly enhancing a flux rope prior to a CME \citep{LiuLJ&al2018, WangW&al2019, Nindos&al2020, James&al2020}. The present event gives a particularly clear indication of flux rope formation by a confined flare because the rope is clearly distinct from the seed rope in the NLFFF model and from the initially formed (first) hot channel. One foot point of the resulting rope is located in a separate sunspot, far away from the initially erupting flux. 

When erupting flux is halted in the corona and remains confined, the addition of current-carrying flux is usually required to produce a CME subsequently. This follows from the finding that a full eruption requires a certain fraction of the active-region flux to reside in the current-carrying, erupting flux \citep{Bobra&al2008, Savcheva&vanBallegooijen2009, SuYN&al2011}. Alternatively, a part of the stabilizing overlying flux can be removed, as conjectured, e.g., in \citet{Panesar&al2015}, but this requires a multipolar source structure and an external agent \cite[see, e.g.,][]{Torok&al2011}, which jointly exist only in a minority of the eruptions. If the flare associated with the confined eruption involves the standard flare reconnection, which can be inferred from the formation of flare ribbons and loops, then the addition of flux is already initiated by the flare reconnection of the confined eruption itself. 

It is important to note that the added flux always winds about the erupted flux. 
Usually, a flux rope of about one turn or slightly higher results. The resulting flux may possess less than one turn in special geometries, when the added flux has very remote foot points or is very highly sheared, such that its legs do not pass over the legs of the initially erupted flux. Nevertheless, the basic structure of a flux rope still results with a twist somewhat lower than one turn but typically not much lower. According to a general definition---a twisted flux tube \cite[][Ch.~2.9.1]{Priest2014}---a flux rope does not require a minimum twist. 

Subsequent reconnection, driven from the photosphere, by the further evolution of the erupted flux, or by another confined eruption at the same or a neighboring section of the PIL, can add further flux to the rope. Eventually, a CME can result. Since many CMEs are preceded by one or several confined eruptions which produce flare ribbons and loops, we conclude that a flux rope exists prior to a significant if not large fraction of all CMEs. A statistical investigation estimating the fraction is in progress.

\subsection{Slow Rise due to Non-equilibrium}\label{ss:slow_rise}

The analyzed event suggests that a confined eruption formed a flux rope in a non-equilibrium state in a range of parameter space that is stable against the torus instability, i.e., at sufficiently low heights, where the decay index of the external field is sub-critical. This is the natural combination of properties for the case of flux rope formation or enhancement by confined eruptions. From the confinement, the decay index at the position of the new or enhanced flux rope must be sub-critical. Since the rope is formed or enhanced by a highly dynamical process operating at coronal time scales, there is no reason to expect that an equilibrium structure typically results. This is principally different from flux rope formation by slow tether-cutting reconnection that is driven from the photosphere. Since the driving photospheric motions have velocities three to four orders of magnitude below the coronal Alfv\'en velocity, the coronal field closely follows an equilibrium sequence in this case. 

A stable non-equilibrium flux rope immediately begins to seek an equilibrium state. This is a relaxation process of the coronal field which involves a reconfiguration at the macroscopic scale of the non-equilibrium structure, i.e., basically an ideal-MHD process. (Magnetic reconnection may be helpful or necessary for the process to reach completion, but is not its driver; rather, reconnection is likely driven by the reconfiguration.) Therefore, coronal velocities are to be expected. At the same time, it is unlikely that the relaxation process operates at velocities comparable to those of an ideal MHD instability, which can reach a significant fraction of several tenths, even more than one half, of the coronal Alfv\'en velocity in the flux rope (which is $\sim$$10^3$ up to several $10^4$~km\,s$^{-1}$); see \citet{Kliem&Torok2006} and, e.g., \citet{Torok&Kliem2005}. Consequently, a stable, non-equilibrium flux rope is a candidate driver for the slow-rise phase which often precedes a solar eruption \cite[e.g.,][]{ZhangJ&Dere2006}. Explaining the typical rise velocities of this phase, a few up to $\sim$$10^2$~km\,s$^{-1}$, has been a puzzle. However, they may naturally match the velocities built up by the relaxation of stable, non-equilibrium coronal flux ropes. The associated reconnection may typically proceed at intermediate rates, resulting in the simultaneous slow rise of the soft X-ray flux. Quite often, the slow-rise phase commences with a confined eruption and includes one or several confined eruptions during its evolution \cite[see, e.g.,][]{ChengX&DingMD2016a, WangW&al2019}.

\subsection{Driven Reconnection due to Non-equilibrium} 

In some events, the reconnection driven by a stable, non-equilibrium flux rope may also be stronger, reaching levels typical of a flare, especially if a strong flux rope is formed by a significant preceding flare. The compound event investigated here shows a spatial and temporal synchronization between the expanding elbows of the hot channel (Figure~\ref{f:sigmoid}(b)--(c)), the transient ribbon extensions (Figures~\ref{f:ribbons}(b)--(c) and \ref{f:aia304+193}(b)), and the impulsive phase of the second confined flare (Figure~\ref{f:lightcurves}), all starting around 01:30~UT and the latter two being of similar duration ($\sim$10~min). Driving of the second confined flare by a further eruption of flux in the core region is unlikely, because the flux would have to overcome the tension force of the re-closed flux in the flare loops of the first confined flare. Therefore, we suggest that the relaxation of the formed non-equilibrium flux rope strongly---if not mainly---drives the reconnection of the second confined flare. 

The expansion of the elbows visualizes the rise of flux in the area of the ribbon extension. The unusually strong bending of the elbows indicates a mainly upward expansion which produces the strong bending in projection. In addition to triggering reconnection locally under the rising elbows, this may also induce Lorentz forces that support the reconnection inflow in the core region, where the flare emissions mainly originate. In particular, the rise of flux in the area of the sigmoid elbows (orange field lines in Figure~\ref{f:flines}) could shear the flux in the reconnection inflow regions such that this flux differentially rotates upward in planes parallel to the vertical current sheet. This would induce a Lorentz force pointing toward the vertical current sheet, analogous to the force that creates a current sheet when a strongly horizontally sheared arcade collapses \citep[e.g.,][]{Mikic&Linker1994}. 

On the other hand, the suggested stable, non-equilibrium flux rope differs principally from strongly ``out-of-equilibrium'' flux ropes used in some numerical experiments \cite[e.g.,][]{Manchester&al2008, JinM&al2017}. Such flux ropes contain significantly more flux than can be held in equilibrium by the ambient field, so that a fast eruption immediately commences, regardless of whether the rope is placed in a torus-stable or torus-unstable height range.

\subsection{Sequences of Eruptions}

A strong role for flare reconnection in the build-up of magnetic flux ropes allows for a range of time sequences of confined and ejective eruptions. On one end of the spectrum, one has one or several isolated confined eruptions preceding a CME. Their associated flares fully decay between the eruptions. Slow tether-cutting reconnection between them must be important in building up further current-carrying flux which is added to the previously erupted flux, such that a further eruption can result. 

On the other end, one has a continuously amplifying eruption, beginning with a slower confined eruption out of which a faster full eruption (a CME) seamlessly evolves. The confined eruption remains slower because the confining overlying flux slows down the rise of the unstable flux. The rate of flare reconnection of the confined eruption also remains smaller, and it will usually begin to decrease when the rising flux is decelerated \citep{QiuJ&al2004, Temmer&al2008, Temmer&al2010}. However, the resulting soft X-ray flux usually begins to decrease with a delay. If the enhancement of the erupted flux by the flare reconnection of the confined eruption creates an unstable flux rope before the associated soft X-ray flare emission begins to decrease, then the onset of the full eruption is accompanied by a second rise of the flare emission out of the confined flare. Usually, this rises faster than the emission of the confined flare. There are many examples of such two-step eruptive behavior, notably the ones on 2010 August~1 \citep{LiuR&al2010}, 2014 September~10 \citep{ChengX&DingMD2016a, ZhouGP&al2016}, 2015 November~04 \citep{WangW&al2017, ChengX&al2018}, 2011 June~21 \citep{ZhouZJ&al2017}, and 2015 June~25 \citep{LiuR&al2019}. 

Compound events like the one investigated here lie between these two extremes. They are characterized by the first, confined eruption falling short of the requirement for a CME by a relatively small amount, which is reached after the associated confined (usually impulsive) flare has begun to decay but has not yet ended completely.

\subsection{Formation Time of Flux Rope vs.\ Hot Channel}

In the case of one or several isolated confined eruptions, a formed hot channel may stay visible between the eruptions or cool down and disappear, depending on the duration of the intervals between the eruptions relative to the cooling time. However, even if the channel becomes invisible as a consequence of cooling, the flux rope remains in place. Nearly the same or a strongly enhanced rope may light up in a subsequent eruption. Therefore, in many cases, a flux rope already exists before a hot channel can be detected.

\subsection{Double-decker Flux Ropes}

Flare reconnection yields a flux rope above a vertical current sheet. Ropes with extended bald-patch sections in their central part do not result. If bald patches exist in non-erupting flux under the vertical current sheet, a so-called double-decker flux rope \citep{LiuR&al2012} is formed. Subsequent tether-cutting reconnection may enhance the lower or the upper rope, depending on where the coronal flux, driven by photospheric motions, reconnects.

\subsection{Double-arc Instability}

The process also naturally explains the observation that hot channels often have a dip when they first light up. The curvature of the dipped field lines provides an additional upward Lorentz force, which can trigger the double-arc instability \citep{Ishiguro&Kusano2017}. However, from the confined nature of many eruptions, one can infer that the additional Lorentz force of the dipped field lines often has a smaller effect on the stability of a flux rope than the magnetic pressure-gradient force of the flux between the rope and the photosphere, which is the driver of the torus instability.

\subsection{Main Contribution to Flux Rope Formation and Eruption Forecast}

The question whether the long-lasting but slow tether-cutting reconnection driven from the photosphere or the short but fast flare reconnection of a confined eruption typically provide the main contribution to flux rope formation prior to CMEs requires further study. This is relevant for the forecast of CMEs. 

Even if the initially erupting (but still confined) flux in the present event is already largely organized in a flux rope, the eventually fully erupting flux is a strongly modified flux rope largely rooted in new foot points. This raises the question how reliably the onset time of eruptions can be predicted from models of the coronal magnetic configuration, even if these are directly derived from magnetogram observations. It may take much less time for a confined eruption to make a region ready for a full eruption than photospheric driving. While this possibility complicates the development of forecasting schemes, it is also clear that it requires the source region to reach the conditions for a confined eruption. Improved understanding of confined eruptions will allow deriving quantitative onset conditions for them in the future, especially in the case that these are due to an ideal MHD instability. The threshold of instability then provides a quantitative onset criterion, as for a CME, with the difference presumably only lying in differences of the magnetic flux overlying the unstable flux rope at some larger height. On the other hand, the magnitude and number of confined eruptions required to reach the onset condition for a CME from a given state of the coronal field are very difficult to determine theoretically. Estimating them will likely require empirical experience derived from a large number of events and careful, data-constrained numerical modeling.

\section{Conclusions}\label{s:conclusions}

We have considered a compound eruptive solar event, focusing on the formation of an EUV hot channel and its implications for the formation of a magnetic flux rope. The observations show that the main (second) hot channel of the event is formed by the first confined flare ($\approx$M1.2, peaking at $\approx$01:33~UT), and further enhanced during the second confined flare (M2.1, peaking at 01:42~UT). The full eruption of the hot channel, commencing at $\approx$01:55~UT, represents the onset of the CME and associated (eruptive) flare (M2.6, peaking at 02:36~UT). The following main conclusions are obtained. 

\begin{enumerate}
\item The strongly sigmoidal main hot channel can only be interpreted as a flux rope. The immediate onset of expansion of its elbows, with the central part remaining stationary until the onset of the CME, indicates that the flux rope was formed in a non-equilibrium state in a height range which is stable against the torus instability. The evolving flux rope reaches the threshold of torus instability at the onset of its full eruption into the CME. 
\\[-20pt] 
\item Therefore, this event demonstrates the formation of a flux rope---by a confined flare---at least 25~min \emph{prior to} the onset of its eruption into a CME. 
\\[-20pt] 
\item The reconnection underlying the second confined flare continues until the onset of the eruptive flare, but at a continually decreasing rate, as evidenced by the decreasing soft X-ray and microwave flux. This is opposite to the suggestion by the reconnection models that the onset of eruptions is due to self-amplifying reconnection without an external agent (such as ideal MHD instability). We conclude that the CME and its associated flare conform to the ideal MHD model for eruptions only.
\\[-20pt] 
\item The formation and enhancement of a torus-stable flux rope by one or several confined flares, prior to the rope's full eruption into a CME, is likely to be a generic process. It is also likely that the flux rope is formed or enhanced in a non-equilibrium state. The immediately commencing evolution of such a flux rope in the search for an equilibrium state is a candidate driver for the slow-rise phase that typically precedes CMEs. However, the rope may also relax to an equilibrium, or may strongly support further flare reconnection which leads to a two-step eruption or a compound event. 
\end{enumerate}
\vspace*{-20pt}

\acknowledgments
We thank the anonymous referee for constructive comments, which helped improving the clarity of the paper and stimulated most of the analysis in Sections~\ref{ss:post_eruption} and \ref{ss:homologous}. 
Part of this work was carried out in the framework of the joint research program at the Institute for Space-Earth Environmental Research (ISEE), Nagoya University; B.K. and J.L. acknowledge the invitation by ISEE to join the program. 
This research has made use of data provided by the \textsl{SDO}/AIA and HMI, \textsl{RHESSI}, \textsl{SOHO}/LASCO, and NoRH and NoRP science teams. 
NoRH is operated by the International Consortium for the Continued Operation of the Nobeyama Radioheliograph (ICCON). ICCON consists of ISEE/Nagoya University, NAOC, KASI, NICT, and GSFC/NASA. 
B.K. and R.L. acknowledge support by the DFG and NSFC through the collaborative grant KL817.8-1/NSFC41761134088. 
B.K. also acknowledges support by NASA under Grants NNX16AH87G, 80NSSC17K0016, 80NSSC19K0082, and 80NSSC19K0860, and support by ISSI, Bern for the International Team ``Decoding the Pre-eruptive Magnetic Configuration of CMEs.'' 
R.L. also acknowledges NSFC support under Grants 41774150 and 11925302. 
J.L. and C.L. are supported by NASA grants 80NSSC18K1705 and 80NSSC18K0673, and NSF grants AGS 1821294 and AGS 1927578. 
S.W. is supported by an AFOSR Windows on the World grant.
S. M. is supported by JSPS KAKENHI Grant Number JP18H01253.

\facilities{SDO (AIA, HMI), RHESSI, SOHO (LASCO), GOES, NoRH, NoRP}

% \clearpage 
\bibliographystyle{aasjournal} 
\bibliography{references_compound}
\end{document}